\pdfoutput=1

\documentclass[11pt]{article}

\usepackage[preprint]{style/acl/acl}

\newif\ifcomment\commenttrue  

\usepackage[a-1b]{pdfx}

\usepackage{framed}
\usepackage{mdwlist}
\usepackage{siunitx}
\usepackage{latexsym}
\usepackage{colortbl}
\usepackage{xcolor}
\usepackage{nicefrac}
\usepackage{booktabs}
\usepackage{fnpct}
\usepackage{amsfonts}
\usepackage[T1]{fontenc}
\usepackage{bold-extra}
\usepackage{amsmath}
\usepackage{amssymb}
\usepackage{bm}
\usepackage{graphicx}
\usepackage{mathtools}
\usepackage{microtype}
\usepackage{multirow}
\usepackage{multicol}
\usepackage{xpatch}
\usepackage{latexsym,comment}
\usepackage[normalem]{ulem}

\newcommand*{\missingreference}{{\Huge \colorbox{red}{?reference?}}}
\newcommand*{\missingcitation}{{\Huge \colorbox{red}{?citation?}}}

\makeatletter
\xpatchcmd{\@setref}{\bfseries}{\missingreference}{}{}
\def\@citex[#1]#2{\leavevmode
    \let\@citea\@empty
    \@cite{\@for\@citeb:=#2\do
        {\@citea\def\@citea{,\penalty\@m\ }%
            \edef\@citeb{\expandafter\@firstofone\@citeb\@empty}%
            \if@filesw\immediate\write\@auxout{\string\citation{\@citeb}}\fi
            \@ifundefined{b@\@citeb}{\hbox{\reset@font\missingcitation}%
                \G@refundefinedtrue
                \@latex@warning
                {Citation `\@citeb' on page \thepage \space undefined}}%
            {\@cite@ofmt{\csname b@\@citeb\endcsname}}}}{#1}}
\makeatother

\newcommand{\gem}[1]{\mbox{\textsc{gem}}}
\newcommand{\abr}[1]{\textsc{#1}}

\newcommand{\hidetext}[1]{}
\newcommand{\ignore}[1]{}

\ifcomment
    \newcommand{\pinaforecomment}[3]{\colorbox{#1}{\parbox{.8\linewidth}{#2: #3}}}

    \newcommand{\prtodo}[1]{\pinaforecomment{lightblue}{pr}{#1}}
    \newcommand{\prtodoi}[1]{\pinaforecomment{lightblue}{pr}{#1}}
\else
    \newcommand{\pinaforecomment}[3]{}
    \newcommand{\prtodo}[1]{}
    \newcommand{\prtodoi}[1]{}
\fi

\newcommand{\smallurl}[1]{ \begin{tiny}\url{#1}\end{tiny}}

\definecolor{lightblue}{HTML}{3cc7ea}
\definecolor{CUgold}{HTML}{CFB87C}
\definecolor{grey}{rgb}{0.95,0.95,0.95}
\definecolor{ceil}{rgb}{0.57, 0.63, 0.81}
\definecolor{UMDred}{HTML}{ed1c24}
\definecolor{UMDyellow}{HTML}{ffc20e}

\newcommand{\tydi}[0]{\abr{t}{\small y}\abr{d}{\small i}\abr{qa}}

\usepackage{times}
\usepackage{latexsym}
\usepackage[T1]{fontenc}
\usepackage[utf8]{inputenc}
\usepackage{microtype}
\usepackage{inconsolata}
\usepackage{graphicx}

\usepackage[english,french]{babel}
\usepackage{ragged2e}
\usepackage{hyperref}
\usepackage{authblk}
\usepackage{multirow}
\usepackage{xcolor}
\usepackage{listings}
\usepackage{booktabs}
\usepackage{xspace}
\usepackage{siunitx}
\usepackage{bbding}
\usepackage{enumitem} 
\setlist{noitemsep}   

\newcommand{\megawika}{MegaWika\xspace}
\newcommand{\wikitext}{Wikitext\xspace}
\newcommand{\mwonehf}{\href{https://huggingface.co/datasets/hltcoe/megawika}{\texttt{hltcoe/megawika}}\xspace}
\newcommand{\mwtwohf}{\href{https://huggingface.co/datasets/jhu-clsp/megawika-2}{\texttt{jhu-clsp/megawika-2}}\xspace}
\newcommand*\samethanks[1][\value{footnote}]{\footnotemark[#1]}

\title{\megawika~2: A More Comprehensive Multilingual Collection of Articles and their Sources}  

\author{\textbf{Samuel Barham}\thanks{denotes equal contribution.}}
\author{\textbf{Chandler May}\samethanks}
\author{\textbf{Benjamin Van Durme}}
\affil{
    Human Language Technology Center of Excellence \\
   Johns Hopkins University \\
   \texttt{samuel.barham@jhuapl.edu \{cmay14,\,vandurme\}@jhu.edu}
}

\begin{document}
\maketitle

\begin{abstract}
We introduce \emph{\megawika~2}, a large, multilingual dataset of Wikipedia articles with their citations and scraped web sources; articles are represented in a rich data structure, and scraped source texts are stored inline with precise character offsets of their citations in the article text.
\megawika~2 is a major upgrade from the original \megawika, spanning six times as many articles and twice as many fully scraped citations.
Both \megawika and \megawika~2 support report generation research;
whereas \megawika also focused on supporting question answering and retrieval applications, \megawika~2 is designed to support fact checking and analyses across time and language.
\end{abstract}

\section{Introduction}
\label{sec:introduction}


Collaborative AI for report generation and other complex tasks is in high demand.  Wikipedia is an excellent resource for these tasks; it contains millions of open-access community-authored reports (articles) with claims backed up by external sources (references).
Wikipedia articles also contain structured data such as tabular and relational data provided in tables and infoboxes.
To our knowledge, existing large-scale Wikipedia datasets do not simultaneously capture rich article structure and fine-grained links between sources and individual sentences or phrases.
A large dataset with those features would enable research on a much broader range of problems in report generation, fact checking, and other tasks.

Meanwhile, as in-context learning becomes ubiquitous across domains and modalities, access to high-quality, up-to-date corpora for pretraining and tuning large language models (LLMs) becomes vital.  Wikipedia is often a backbone for such data, as it contains billions of tokens spanning hundreds of languages.  Extracting and scraping sources, linking them to specific claims, and capturing rich article structure would provide more data supporting a larger variety of training and tuning tasks.

The original \megawika dataset consists of 13 million Wikipedia articles spanning 71 million passage-source pairs---30 million with scraped source text---across 50 languages.  \megawika broke ground as a large-scale and highly multilingual report-source dataset~\cite{barham2023megawika}.  We introduce \emph{\megawika~2}, a collection of 77 million articles spanning 172 million citations---63 million with scraped source text---across those same 50 languages.  \megawika~2 improves on \megawika (\emph{\megawika~1}) by providing:
\begin{itemize}
    \item more articles, citations, and text;
    \item sentence-segmented paragraphs;
    \item web and non-web citations (books, etc.) with text offsets;
    \item source text quality estimates;
    \item article elements like tables, infoboxes, and math/code blocks extracted;
    \item improved translations using NLLB; and
    \item enrichments like cross-lingual article links and first/last revision dates.
\end{itemize}
We release \megawika~2 with the aim of accelerating multilingual report generation and related research while facilitating LLM training and tuning.\footnote{
    \mwtwohf
}




\begin{figure*}[ht]
   \centering
    \includegraphics[width=1.0\textwidth]{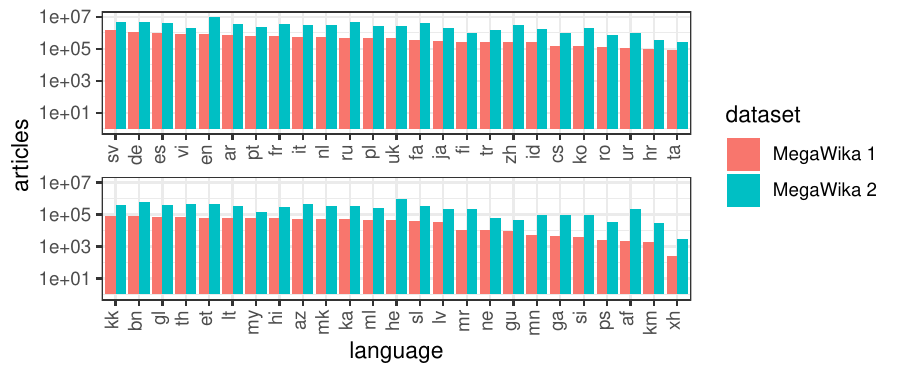}
    \caption{Number of articles (log scale) per language in \megawika~1 and \megawika~2.  Languages are represented by their two-letter Wikipedia project codes (equivalent to ISO 639-1 language codes for these 50 languages) and are ordered by their number of articles in \megawika~1.}
    \label{fig:articles}
\end{figure*}

\section{What's Changed}
\label{sec:whats-changed}

\megawika~2 improves on \megawika~1 in data quality and quantity.  \megawika~2 captures rich article structure, representing citations in the context of the article they appear in; it includes text quality estimates for articles' sources; it provides improved English translations of article text; it contains many more articles, citations, and scraped sources; and it adds data enrichments.

\paragraph{Improved Article Representation}

\megawika~1 focused primarily on the relationship between individual article passages and the web sources they cite, which it represented as a list of passage-source pairs (one source per passage).  \megawika~1 also provided the full text for each article, but the passages and citations were not represented explicitly in that broader context.

By contrast, \megawika~2 captures the structure of the article, detecting paragraph boundaries and exposing headings, paragraphs, tables, infoboxes, and more.  Each paragraph contains a list of sentences, each of which contains a list of citations and citation-needed markers with corresponding character offsets.  Sources are thus represented in the context of the surrounding article and their precise locations in the text are preserved.  Moreover, all citations are captured, not just citations containing URLs (however, only web citations' content is scraped and stored in the data set).  The \megawika~2 dataset is described in further detail in Section~\ref{sec:dataset}.

\paragraph{Source Text Quality Estimates}

Extracted source text in \megawika~1 was often low-quality, making the sources difficult to use in an automated fashion in downstream research.  Many of the circumstances underlying these low-quality sources are effectively unavoidable, including paywalls and CAPTCHAs, other access-restricted pages, and deleted or renamed pages.

To mitigate this issue in \megawika~2, we provide five-level ordinal text quality estimates for all scraped sources.  Our text quality estimates are tuned for precision, enabling users to reliably filter out lower-quality source text as desired.

\paragraph{Improved Translation}

\megawika~1 used M2M100 to translate citation-terminated passages into English, using Google Translate for lower-resource languages where M2M100's performance was unsatisfactory.
\megawika~2 uses NLLB-200 (600 million parameters) to translate all sentences and headings to English, providing improved translation performance overall, as discussed in Section~\ref{sec:translation-analysis}.
Additionally, \megawika~2 contains over five times as many sentences (and headings) as \megawika~1, so the workload is now much larger.
Accessing a pool of 16 shared NVIDIA H100 GPUs on the Johns Hopkins University Data Science and AI Institute (DSAI) grid, we were able to translate the one billion sentences and headings of \megawika~2 in just over two weeks, whereas translating the 180 million sentences in \megawika~1 took months.

\paragraph{Increased Quantity}

\megawika~2 has six times as many articles, over five times as much text---one billion sentences and headings compared to 180 million sentences---and twice as many citations with text extracted as \megawika~1.  See Figure~\ref{fig:articles} for a breakdown of the number of articles in each data set by language, and see Section~\ref{sec:citation-analysis} for a detailed analysis of citations.


\section{Data Processing}
\label{sec:data-processing}


\megawika~2 data processing is divided into four pipelines, allowing us to run data processing steps with different hardware requirements on different systems.
 
\begin{enumerate}
    \item First, the \emph{extraction pipeline} downloads a Wikipedia dump, extracts the articles, organizes the collection into \num{1000}-article chunk files, parses the \emph{\wikitext} code (Wikipedia markup code) of each article, identifies citations, and, for citations that point to a URL, downloads the HTML code or plain text at that URL and extracts text content from it.  This pipeline is CPU- and memory-intensive.
    \item Next, the \emph{translation pipeline} extracts headings and sentences from articles in a chunk, translates them to English, and reinserts them into the collection.  This pipeline is GPU-intensive.
    \item The \emph{source quality pipeline} extracts source texts, estimates source text quality, and inserts the estimates back into the collection.  It is also GPU-intensive.
    \item Finally, the \emph{enrichment pipeline}
    extracts editor-provided quotes from citations,
    fetches cross-lingual links between each article and parallel articles in other Wikipedia projects (languages) as well as the creation date of each article.  This pipeline's bottleneck is the Wikipedia Action API rate limit.
\end{enumerate}

We implement these pipelines using \href{https://luigi.readthedocs.io/}{Luigi}, a Python framework that facilitates designing and running complex data pipelines on a single machine (with tools for invoking external systems like Spark).\footnote{
    Luigi is simpler to set up and use than cloud-ready frameworks like Airflow while providing more flexibility than automation tools like Doit, making it a good middle ground for our computational needs and the research-oriented platforms we have access to.
}

\subsection{Extraction}
\label{sec:extraction}
The extraction pipeline consists of four main steps: downloading the Wikipedia dump, parsing the dump XML and organizing articles into chunks, parsing articles' \wikitext, and scraping sources for web citations.  The \wikitext parsing and source scraping steps are parallelized across chunks.

\paragraph{Dump Download}
The dump download step downloads the Wikipedia dump for a specified language and dump date from \href{https://dumps.wikimedia.org/backup-index.html}{the Wikimedia download server}.
Since old dumps are regularly deleted from the server, the dump download must happen relatively soon (within a couple of months, at the time of writing) after dump creation.

\paragraph{Chunking}
The chunking step extracts articles' \wikitext and metadata from a Wikipedia XML dump and writes out ``chunk'' files of \num{1000} articles (fewer for the final chunk), one per line, in JSON-lines format.  These chunk files form the basis of our intermediate data representation; \num{1000} articles is a reasonable unit of work for most steps.


In this step, we also filter out some special articles like redirects and category pages.\footnote{
    Specifically, we filter out pages that contain \texttt{\#redirect} or \texttt{\{\{website-stub\}\}} (both case-insensitive) in their \wikitext and pages that contain \texttt{Category:} (case-sensitive) in their title.
}
These pages tend to have few or no citations.

\paragraph{\wikitext Parsing}
The \wikitext parsing step parses the articles' \wikitext code into a list of elements representing the article structure, including sentence-segmenting paragraphs and other blocks of text.  It also creates the initial list of excerpts with citations based on those article elements.

\wikitext parsing is nontrivial~\cite{dohrn2011design,siebenmann2011notparsing} and there are many third-party software libraries offering approximate solutions.
We use \href{https://mwparserfromhell.readthedocs.io/}{MediaWiki Parser from Hell} to parse the \wikitext and then postprocess it using a custom procedure.  This postprocessing includes, among other things, text cleaning using \href{https://ftfy.readthedocs.io/}{ftfy}~\cite{speer-2019-ftfy} and \href{https://beautiful-soup-4.readthedocs.io/}{BeautifulSoup}, filtering out file and media wikilinks (links to other Wikipedia pages), detecting paragraph boundaries, sentence-segmenting paragraphs, and resolving links between citations.

Substeps like filtering out file and media wikilinks and detecting infoboxes are surprisingly complex.  This complexity arises from the fact that each Wikipedia project (language) is independent and has its own conventions and resources (to an extent).  Concretely, file and image page titles in English are often prefixed with \texttt{File:} and \texttt{Image:} respectively, but these prefixes are different in other languages, and there will often be multiple prefixes for the same kind of object (most notably, many languages have \texttt{File:} and/or \texttt{Image:} prefixes along with in-language prefixes).  Additionally, articles can link to pages in other Wikipedia projects (languages) by prepending further project-specific prefixes.

Infoboxes are floating boxes rendered on the sides of some Wikipedia pages, for example, a brief list of facts about the subject of a biographical page.  Infoboxes are encoded as \emph{templates}, a special kind of page that is parametrized and can be \emph{transcluded}---basically, included, but with parameter substitution---in other pages.  Simple templates do things like convert a quantity to different units and format the results, or present a phrase alongside with its translation in a different language.  There is a \texttt{Template:} page title prefix for templates in English; other languages have their own prefixes, and templates can be transcluded across Wikipedia projects as well.  Moreover, there is no standardization of infobox templates; most English infobox template titles start with \texttt{Template:Infobox}, but there are many exceptions to that rule, and other Wikipedia projects have other conventions.

We approach these issues empirically.  Offline, we use heuristics, cross-lingual page links, and manual filtering to find file, media, and template prefixes and lists of infoboxes in each Wikipedia projects' articles' \wikitext.\footnote{
    The list of infoboxes especially is subject to change over time; however, we expect that the most commonly used infoboxes will be stable, as changing or removing them could cause a large number of disruptions across Wikipedia.  Widely used template pages have notices warning potential editors that the templates are used in a large number of pages and any changes should be tested in a sandbox area first; and the most prominent are protected and editable only by trusted users.
}
Then, when running the extraction pipeline, we provide those lists as configuration data to the \wikitext parsing step.

\paragraph{Source Scraping}
In the source scraping step, we attempt to download the source content at the URL for each web citation, limiting each download to a timeout of ten seconds and skipping it if the decoded content size exceeds one million characters (approximately 1~MB for most English-language content; larger for content in languages not using the Latin alphabet).  We then attempt to extract text from the downloaded content using \href{https://trafilatura.readthedocs.io/}{Trafilatura}~\cite{barbaresi-2021-trafilatura}, excluding HTML comments but including tables and links.  The output is Markdown, but often contains little Markdown syntax beyond a few hyperlinks.  Finally, we filter out text with less than 100 whitespace-separated tokens; such little extracted text often results from HTTP 404 pages, paywalled pages, and other unusable pages.\footnote{
    This filter is problematic on languages like Chinese that don't use whitespace characters to separate words; we plan to revisit it in future work.
}\footnote{
    By comparison, \megawika~1 omits source text with fewer than 500 characters.
}

\subsection{Translation}
The translation pipeline consists of three main steps:  Extracting sentences and headings from each article, translating sentences and headings to English, and inserting the translations into each article.

\paragraph{Sentence/Heading Extraction}  The first step of this pipeline extracts sentences and headings from each article, storing them in a more lightweight JSON format.  Separating this step from translation enables faster data transfer from the JHU IDIES SciServer platform where the extraction pipeline is run to the DSAI grid where we have more GPU availability, and it reduces GPU job (de)serialization overhead.

\paragraph{Sentence/Heading Translation}  The second step of this pipeline launches a Slurm grid job that allocates a GPU and translates the sentences and headings to English using NLLB-200 600M with 8-bit distillation~\cite{nllbteam2022languageleftbehindscaling}.

\paragraph{Sentence/Heading Insertion}  The third and final step of this pipeline inserts the translated sentences and headings back into the corresponding articles.  Separating this step from translation speeds up data transfers between our computing platforms and reduces (de)serialization overhead, as for sentence/heading extraction.

\subsection{Source Quality}
The source quality pipeline is structured similarly to the translation pipeline; it extracts source texts instead of headings, and predicts source text quality instead of English translations.  Source quality estimates are produced by the ordinal regression model described in Section~\ref{sec:source-text-quality-analysis}.  Due to this pipeline's similarity to the translation pipeline, we do not describe it in further detail.

\subsection{Enrichment}
The enrichment pipeline consists of two main steps:
fetching cross-lingual links between articles from \href{https://www.mediawiki.org/wiki/API}{the Wikipedia Action API} and fetching each article's creation date.
Rate-limiting on the Action API requires the latter two steps to be run serially (one chunk at a time).
The enrichment pipeline also currently includes a step that extracts editor-provided \emph{source snippets} (quotes) from citations.\footnote{
    This step is essentially a parsing operation, and we plan to incorporate it in \wikitext parsing in the extraction pipeline in the future.
}

\paragraph{Source Snippet Extraction}
The source snippet extraction step checks each citation's \wikitext for an author-provided quote from the source (in principle, a specific quote backing up the claim in the article), extracts one if it exists, and attaches it to the citation data structure.\footnote{
    The quote, or snippet, is assumed to be stored in a template parameter named \texttt{quote}.  As templates vary across Wikipedia projects, this is an area that could be improved in future work.
}

\paragraph{Cross-Lingual Links Retrieval}
The cross-lingual links retrieval step uses the Wikipedia Action API to retrieve, for each article, a set of \emph{cross-lingual links}, or links to articles on the same topic in other Wikipedia projects (languages).  We do not resolve article redirects\footnote{
    Redirects are special pages that automatically redirect to other articles; they are largely invisible to users, except when a user follows a link to a redirect page and sees a small notice at the top of the displayed article explaining that they were redirected.  Redirects can be created manually and are also created automatically when a page is moved (or renamed) from one title to another; the article at the old title is automatically converted into a redirect.
}
when fetching cross-lingual links.  Each cross-lingual link consists of a Wikipedia project code and an article title in that project.  The set of cross-lingual links is attached to the article data structure.

\paragraph{Creation Date Retrieval}
The creation date retrieval step uses the Wikipedia Action API to retrieve each article's creation date, the date (more precisely, datetime) of the first revision, and attach it to the article.  We do not resolve article redirects when fetching cross-lingual links.  Creation dates are complicated by redirects; a regular article can be converted to a redirect (or vice-versa), and a cursory investigation suggests that creation dates reflect the date of an article's most conversion to/from a redirect, but more work is needed to map out the full behavior.

\subsection{Incremental Updates}

To enable basic temporal studies at one-month granularity, we have implemented incremental updating in each pipeline.

Wikipedia dumps are released twice a month, usually on the first and twentieth day of the month. 
After a pipeline finishes its first run on a full Wikipedia dump, it can be run in an incremental fashion on subsequent dumps, processing only those articles that have changed or been added since the previous dump.  We call this option the \emph{delta} mode.  The extraction pipeline still downloads and stores the full Wikipedia dump in delta mode, but only new and changed articles are processed in subsequent steps.

\section{Dataset}
\label{sec:dataset}


The \megawika~2 data set consists of a set of directories, one per language (Wikipedia project), each of which contains a set of JSON-lines (\texttt{.jsonl}) files.  Each file represents a chunk of up to \num{1000} articles, one article per line.  Articles are ordered by their location in the corresponding Wikipedia dump.\footnote{
    Overall, articles earlier in a dump have earlier creation dates (first revision dates).  However, creation dates are affected by article moves (renames) and conversion of regular articles to/from redirect pages in seemingly complex ways.  At the time of writing, we do not have a comprehensive model of how articles are ordered.
}
Only the last chunk for each language can contain fewer than \num{1000} articles.

Each article is a JSON object containing the article title, the article \wikitext code, the aggregated text extracted from the article, a list of page elements, and a list of ``excerpts with citations'' (defined later).  Article elements include headings, paragraphs, infoboxes, tables, and other block elements.  Paragraphs, in turn, contain a list of sentences, and each sentence contains the sentence text as well as a list of citations with character offsets into that text. 
 Where they are detectable in the \wikitext, citations also include short author-provided ``snippets,'' or quotes, from the citation source.

The \emph{excerpts with citations} list is a postprocessed subset of the information in the elements list.
It contains a list of \emph{excerpts}, where we define an excerpt as a span of three consecutive sentences within a paragraph (or fewer than three, if at the beginning of the paragraph) such that the last sentence contains at least one citation.  Each excerpt is paired with the citations from its final sentence.  Excerpts' text may overlap.\footnote{
    The excerpts with citations are provided to facilitate comparison with \megawika~1, which consists of passage-source pairs.  We use the term ``excerpt'' instead of ``passages'' in \megawika~2 to better differentiate passages (excerpts) from paragraphs.  Note that excerpts in \megawika~2 may overlap while excerpts in \megawika~1 do not.
}

The \megawika~2 data structure is described in further detail in Appendix~\ref{appendix:schema}.  The dataset is available on HuggingFace at \mwtwohf; it will be provided without scraped source content but with source rehydration (scraping) code.

In the following subsections, we provide detailed analyses of the citations and translations in \megawika~2.  These analyses were facilitated by the \emph{GNU parallel} parallel computing tool~\cite{tange2024gnu}.

\subsection{Summary Statistics}

\begin{table}[ht]
    \centering
    \begin{tabular}{lrr}
        \toprule
        \textbf{Type} & \textbf{All Langs} & \textbf{English} \\\midrule
        Articles & 77~M & 10~M \\
        Headings & 183~M & 33~M \\
        Paragraphs & 336~M & 68~M \\
        Sentences & 1~B & 228~M \\
        Citations & 237~M & 74~M \\
        Web Citations & 172~M & 57~M \\
        Sources & 63~M & 22~M \\
        \bottomrule
    \end{tabular}
    \caption{Count of each element type in all of \megawika~2 and (separately) in the English subset of \megawika~2.  \textbf{Sources} refers to the web citations that have text extracted (scraped).  M means million, B means billion.}
    \label{tab:summary-statistics}
\end{table}

The 50 languages in \megawika (hence \megawika~2) were selected according to the scale of their Wikipedia projects while ensuring coverage of a diverse set of language families~\cite{barham2023megawika}.
Across those 50 languages, \megawika~2 contains 77 million articles, 10 million of which come from English Wikipedia.  Additional summary statistics are provided in Table~\ref{tab:summary-statistics}, and Figure~\ref{fig:articles} shows the number of articles in each dataset grouped by language.

Of the 87 million unique URLs detected across \megawika~2 web citations, without normalization, only 1.4 million are also present in FineWeb-Edu.  This relatively small intersection suggests that our collection of web citation URLs from Wikipedia---about 25\% of which can (or could recently) be scraped---is a novel contribution in itself.

\subsection{Citation Analysis}
\label{sec:citation-analysis}

We now compare citations in \megawika~1 and \megawika~2 quantitatively.

\megawika~1 reports 13 million articles and 71 million sources~\cite{barham2023megawika}.  However, in the version of the dataset we downloaded from HuggingFace (\mwonehf), we found 13 million articles and 63 million sources (30 million of which have non-empty source text extracted).  In this section, we use statistics computed from this copy of \megawika~1.

Additionally, the English subset of our copy of \megawika~1 has three million web citations, and all of them have non-empty text extracted.  This idiosyncrasy suggests web citations with empty text have been filtered out of this subset, as a 100\% download and extraction rate across millions of web pages linked from Wikipedia is virtually impossible.

\begin{figure*}[ht]
   \centering
    \includegraphics[width=1.0\textwidth]{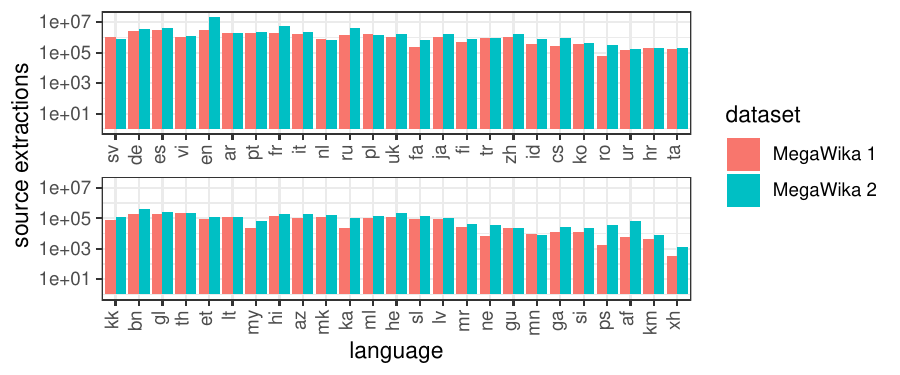}
    \caption{Number of sources extracted (with non-empty extracted text) (log scale) per language in \megawika~1 and \megawika~2.}
    \label{fig:source-extractions}
\end{figure*}

\paragraph{Sources Extracted per Article}
As reported in Section~\ref{sec:introduction}, \megawika~2 has more web citations detected and sources extracted (with non-empty text) than \megawika~1.  A breakdown of sources extracted by language is presented in Figure~\ref{fig:source-extractions}.  However, the number of source extractions per article is much lower in \megawika~2 than in \megawika~1 for all but two languages---Burmese (my) and Pashto (ps)---as shown in Figure~\ref{fig:source-extractions-per-article} in Appendix~\ref{appendix:citation-analysis-extra}.

To understand why fewer sources are extracted per article in \megawika~2, we compare articles only in \megawika~1, articles only in \megawika~2, and articles in both datasets (in the intersection).

\begin{figure*}[ht]
   \centering
    \includegraphics[width=1.0\textwidth]{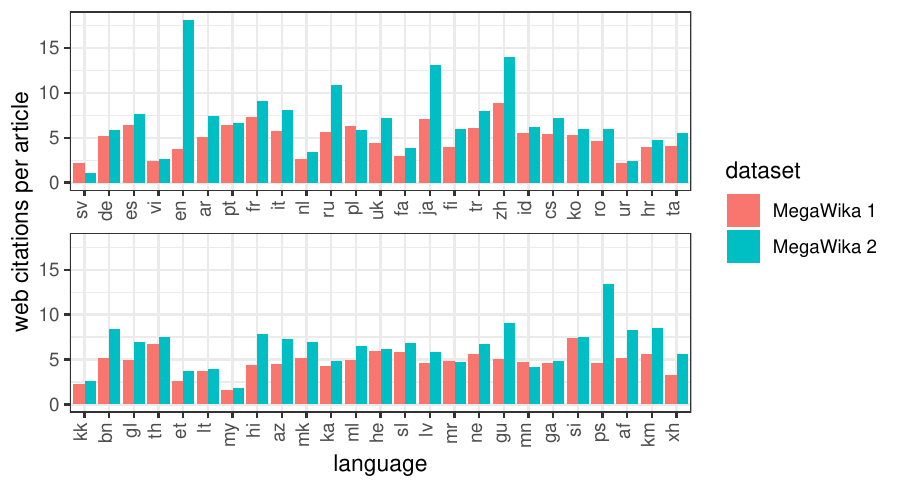}
    \caption{Average number of web citations per article for article titles appearing in both \megawika~1 and \megawika~2.}
    \label{fig:mw1-and-mw2-sources-per-article}
\end{figure*}

For all 50 languages, the majority of article titles found in \megawika~1 \emph{or} \megawika~2 (in the union) appear only \megawika~2, as shown in Figure~\ref{fig:mw1-mw2-venn-diagrams} in Appendix~\ref{appendix:citation-analysis-extra}.  
Virtually all of the remaining article titles---those appearing in \megawika~1---appear in the intersection of \megawika~1 and \megawika~2.
Among article titles appearing in the intersection, for all but four languages---Marathi (mr), Mongolian (mn), Polish (pl), and Swedish (sv)---we detected more sources per article in \megawika~2 than in \megawika~1, as shown in Figure~\ref{fig:mw1-and-mw2-sources-per-article}.

\begin{figure*}[ht]
   \centering
    \includegraphics[width=1.0\textwidth]{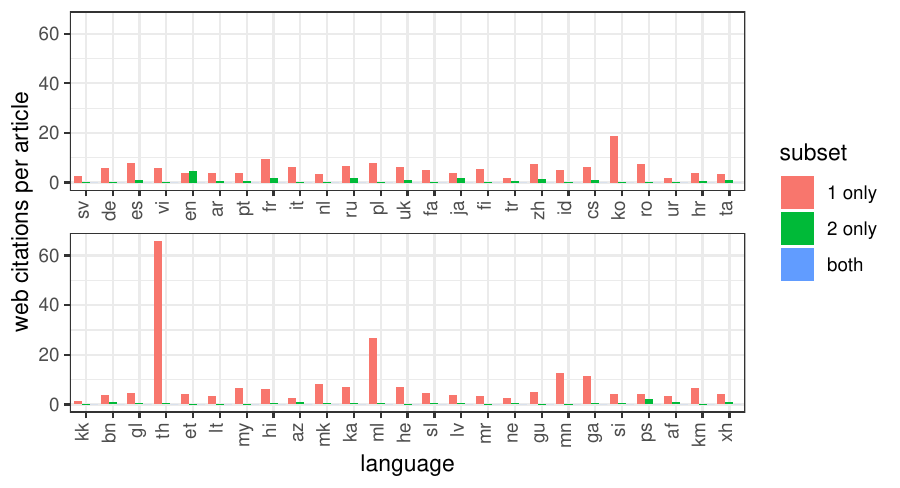}
    \caption{Average number of web citations per article for titles only in \megawika~1 (red bars) and for titles only in \megawika~2 (green bars).}
    \label{fig:mw2-xor-mw1-sources-per-article}
\end{figure*}

Among the titles appearing in \megawika~2 but not \megawika~1 (which make up the majority of all titles in the union), we detected very few sources per article, as shown in Figure~\ref{fig:mw2-xor-mw1-sources-per-article}.
For all but one language---English (en)---the mean number of sources detected per article among the (relatively few) articles appearing in \megawika~1 but not \megawika~2 was much higher.

A cursory qualitative investigation suggests that titles only in \megawika~1 include many pages that were been deleted or renamed before \megawika~2 data collection.  On the other hand, we found that titles only in \megawika~2 include:
\begin{itemize}
    \item special page types like files, templates, and categories, which tend to have few or no citations;
    \item short articles and stub pages (perhaps indicating newly created articles) which have fewer citations due to their shorter text;
    \item pages that have mostly list-like or tabular content (perhaps reproducing information retrieved from a single source), like discographies and competition results; and
    \item pages that had been renamed since \megawika~1 data collection.
\end{itemize}

That is, titles only appearing in \megawika~1 include many articles that were renamed or deleted before \megawika~2 data collection, and we hypothesize that many of these were established articles with high citation counts.  Titles only appearing in \megawika~2 include renamed articles as well, but those are outnumbered by a variety of new and/or special articles that tend to have low citation counts.  Thus, overall, \megawika~2 has a lower number of citations per article than \megawika~1.

\paragraph{Source Extraction Rate}
We also find that the source extraction rates (the fraction of web citations with sources extracted with non-empty text) are generally lower in \megawika~2 than in \megawika~1.  A breakdown of source extraction rates is presented in Figure~\ref{fig:source-extraction-rates} in Appendix~\ref{appendix:citation-analysis-extra}.

We initially hypothesized that potentially new or updated rate limits from \href{https://web.archive.org/}{web.archive.org} (the Wayback Machine), which archives the current versions of web pages on demand and is frequently referenced on Wikipedia,\footnote{
    Wikipedia encourages archiving sources in its \href{https://en.wikipedia.org/wiki/Wikipedia:Citing_sources\#Preventing_and_repairing_dead_links}{author guide}.
}
might be the cause of the lower source extraction rates in \megawika~2.  However, we found that filtering to non-web.archive.org sources did not fully explain the drop in source extraction.  As shown in Figure~\ref{fig:non-web-archive-source-extraction-rate} in Appendix~\ref{appendix:citation-analysis-extra}, for most languages, the source extraction rate on non-web.archive.org sources is still lower in \megawika~2, even if only slightly.

To uncover the cause of the drop, we perform a small qualitative study of source extraction failures.
Since web.archive.org rate-limits us and produces many predictable scrape failures as a result, we draw 30 uniform-at-random samples of non-web.archive.org pairs of article titles and source URLs that appear in both \megawika~1 and \megawika~2 (across all languages) and whose sources were successfully extracted in \megawika~1 but not in \megawika~2.  We then attempt to download and extract those sources' content again using the \megawika~2 source scraping code.  Finally, the second author manually categorized the results, tallying the number of scrapes in each category.


\begin{table}[ht]
    \centering
    \begin{tabular}{lr}
        \toprule
        \textbf{Category}         & \textbf{Count} \\\midrule
        Success                   & 2 \\\midrule
        \textit{Download error}   &   \\
        \ -- Max retries exceeded & 6 \\
        \ -- Read timeout         & 1 \\\midrule
        \textit{Extraction error} &   \\
        \ -- HTTP 403 (forbidden) & 7 \\
        \ -- HTTP 404 (not found) & 9 \\
        \ -- HTML skeleton only\footnote{This category includes pseudo-404 responses in which a body displaying ``page not found'' (or similar) is sent with a success (200--209) status code.} & 3 \\
        \ -- HTML body w/ few words\footnote{Recall we do not store extracted source text that does not meet a minimum (whitespace-separated) word count; pages with few such ``words'' often contain ``page not found'' text, paywalls, or other kinds of irrelevant content.} & 2 \\
        \bottomrule
    \end{tabular}
    \caption{Categorization of re-scrapes of a uniform sample of web citations whose sources were successfully downloaded and extracted in \megawika~1 but not in \megawika~2.}
    \label{tab:source-rescrape-categories}
\end{table}

The results of this analysis are presented in Table~\ref{tab:source-rescrape-categories}.  All but two response categories, successful re-scrapes and HTML body content with few words, arise from factors largely outside our control.  Thus, we might expect to improve the source extraction rate by an upper bound of about 15\% by making our source scraping approach more robust.

We performed this experiment weeks after \megawika~2 source scraping, so we argue the two successful re-scrapes represent virtually all sources in this sample that we might have successfully extracted by implementing additional retries over time.

The two re-scrapes that produced HTML content with too few words were a product listing with little text content and a page of largely image content appearing to promote a mobile app.  We argue that inclusion of such sources in \megawika~2 would not be worth the concomitant inclusion of degenerate cases like HTTP 404 ``not found'' pages returned with successful status codes, paywalls, navigation pages with little text content, and pages whose content is dynamically generated by minified JavaScript.

These findings suggest that sources cited by Wikipedia are gradually (at least partially) becoming inaccessible.  This phenomenon is known as \emph{link rot}, and Wikipedia has \href{https://en.wikipedia.org/wiki/Wikipedia:Link_rot}{a special page} describing it and steps that have been (and can be) taken to mitigate it.

According to that page, Since 2015, Wikipedia has used bots to automatically archive sources (and other pages) linked in the English and other-language Wikipedias and introduced a bot that automatically detects bad links and attempts to replace them with archive versions.  Wikipedia also encourages authors to manually archive sources, and many citation templates provide facilities to copy relevant source quotes directly into the citation (the basis of our source snippets).  However, the bots cannot archive some links due to technical barriers, and the bot that proactively archives sources only checks page \emph{edits}, not the existing page content.  These limitations may be why many \megawika~1 sources are now inaccessible.

Additional plots for both the citation per article analysis and the source extraction rate analysis are provided in Appendix~\ref{appendix:citation-analysis-extra}.




\subsection{Source Text Quality}
\label{sec:source-text-quality-analysis}

A key challenge in making \megawika~2 useful to other researchers is ensuring the quality of scraped web sources.  A substantial fraction of the scraped sources consist of ``page not found'' or ``forbidden'' messages (returned with HTTP 200 status codes), irrelevant content like ads, content that is very difficult to parse into plain text, or content that is otherwise unsuitable for use as source text.  Rather than filtering sources by quality at an arbitrary threshold, we provide a predicted ordinal source quality annotation.

To produce these annotations, we developed a source quality prediction model that could be deployed in the \megawika pipeline.  This process consisted of manual and LLM-assisted exploration of source quality variation, several rounds of human annotation of source quality and rubric development, LLM generation of silver source quality labels, and source quality prediction model training and selection.

\subsubsection{Gold Human Annotation}

To understand the failure modes of source scraping, we first manually reviewed a small sample of scraped source texts and also asked an LLM to review and categorize a larger sample of texts.  Together with the LLM, we identified a number of cases, including: HTTP 404 pages, paywalls, and CAPTCHAs; ads and lists of ``related stories;'' pages with lingering encoding errors; pages with mangled or incomplete table formatting; pages with apparently incomplete text; and pages largely containing metadata instead of the likely intended source content.  We used this exploratory analysis to inform the later development of our source quality taxonomy.

The first two authors then labeled 100 examples of scraped source texts.  These 100 examples were chosen by randomly sampling source texts from \megawika~2 (stratifying by language), truncating them to \num{5000} characters, and translating them to English using the Azure text translation service.

The annotation interface displayed the original and translated source text and provided a 100-point slider for rating source quality.  It also showed a five-point ordinal label corresponding to the value of the slider, computed using equal-width bins.  Initially, no class descriptions were provided; annotations were only instructed that 1 corresponded to the lowest quality and 5 to the highest.  This approach allowed us to first capture our intuitions about source quality before committing to a more structured rubric.

Following the initial round of annotation, we identified and discussed the 25 sources with highest inter-annotator disagreement as measured by the absolute difference.  In these discussions, we sought to: (1) develop a shared understanding of levels of quality/modes of failure in the scraped source texts, (2) identify appropriate potential granularities for the ordinal source quality scale, and (3) attempt to resolve any major systematic disagreements in the quality of scraped source text.  We decided that either a three-point or five-point scale could work well, and we concurred on a single label of each granularity for most of the initial disagreements; the remaining cases were generally intermediate-quality sources on which we only disagreed by small amounts.  We then devised a rubric for a five-point scale:

\begin{itemize}
    \item \textbf{Class 1}: \textit{Not relevant to the intended source material; may be text from a paywall, HTTP 404 or other error message, CAPTCHA page, site navigation menu, promotional content, link farm, etc.}
    \item \textbf{Class 2}: \textit{Unlikely to be relevant to the intended source material and interpretable; may be a list of unrelated article excerpts or a table that was mangled by extraction.}
    \item \textbf{Class 3}: \textit{May or may not represent the intended source text; often a book abstract or website ``about'' page that suggests the scraper was redirected away from the desired content.}
    \item \textbf{Class 4}: \textit{Represents the intended source text with some readability or formatting issues; may include repeated sections, distracting artifacts, or embedded irrelevant text.}
    \item \textbf{Class 5}: \textit{Represents the intended source material with minimal readability or quality issues; generally coherent and usable, though may contain a few minor markdown links or extraneous formatting.}
\end{itemize}

Classes 1 and 2 represent irrelevant and/or poorly-formatted text, class 3 represents text that may or may not faithfully reflect the intended source material, and classes 4 and 5 represent higher-quality text that is likely to capture the intended source information and be sufficiently well-formatted for use in downstream tasks.

With this initial rubric developed, we added the class label descriptions to the annotation interface, re-annotated the first 100 examples, and annotated 100 additional examples.  We used this set of 200 source quality labels as gold data to develop and evaluate a larger LLM annotation dataset.

\subsubsection{Silver LLM Annotation}

Given the time and cost associated with large-scale human annotation, we next used LLMs to automatically annotate a larger set of scraped sources with quality labels for model training.  We evaluated both five-point and three-point LLM output scales to verify that LLMs could consistently recover the more nuanced five-point categorization.

\paragraph{Prompt Engineering}
We first developed a basic prompt format tailored to GPT-4o via Azure OpenAI endpoints. The design goal was twofold: to ensure a clean, parseable output format with clearly separated scores and justifications (the latter constructed to aid diagnosis during prompt development), and to maximize correlation with human annotations. The final prompt, provided in Appendix~\ref{appendix:silver-annotation-prompt}, adopts the perspective of a human analyst vetting a source for inclusion in a written article.  It presents the first $N$ characters of a scraped source, the originating URL, and the quality rubric.  The model is then asked to assign an ordinal score (1--5 or 1--3) reflecting the degree to which the content is usable and relevant.

In an attempt to optimize the silver annotations' correlation with human gold annotations, we conducted a small grid search over several annotation dimensions:
\begin{itemize}
    \item \textbf{Input language:} using the original source text in its native language, versus using English translations obtained via Azure’s translation API.
    \item \textbf{Output scale:} predicting coarse ordinal class labels (1--5 or 1--3, as described above) versus predicting scores on a 1--100 scale, which we refer to as ``continuous'' (though they are really fine-grained ordinal labels). The mapping from this continuous scale to the three- and five-point ordinal scales was included in the prompt to aid model comprehension. Each ordinal class was evenly spaced across the continuous range; for the five-point scale, values 1--20 map to class 1, 21--40 to class 2, etc.
    \item \textbf{Levels:} the LLM and human annotations were mapped to an ordinal scale with three or five levels before computing correlations.
\end{itemize}

We evaluated the resulting annotations against our gold standard using Pearson, Spearman, and Kendall correlation, as reported in Table~\ref{tab:llm-human-correlation}; ``continuous'' outputs were mapped to a three- or five-point ordinal scale before computing correlation.  We find that differences between original sources and automatically translated sources, and between continuous and ordinal scales, are small.  Using a three-point scale sometimes yields a substantial increase in correlation over a five-point scale.

\begin{table*}[ht]
\centering
\caption{Pearson, Spearman, and Kendall correlations between LLM silver annotations and human gold scores, using both three- and five-point ordinal scales. All correlations were computed over ordinal labels. For ``continuous'' LLM outputs (1--100 scale), values were mapped to ordinal categories using even intervals, as described in the text. This mapping was explicitly included in the LLM prompt, just as the UI shown to human annotators displayed the correspondence between continuous and categorical labels that would be used for gold star annotations.}
\label{tab:llm-human-correlation}
\begin{tabular}{lllrrr}
\toprule
\textbf{Output Scale} & \textbf{Levels} & \textbf{Language} & \textbf{Pearson} & \textbf{Spearman} & \textbf{Kendall} \\
\midrule
ordinal & 5 & translated & 0.83 & 0.69 & 0.64 \\
continuous & 5 & translated & 0.85 & 0.72 & 0.67 \\
ordinal & 5 & original & 0.84 & 0.68 & 0.63 \\
continuous & 5 & original & 0.80 & 0.71 & 0.66 \\
\midrule
ordinal & 3 & translated & 0.85 & 0.80 & 0.77 \\
continuous & 3 & translated & 0.82 & 0.75 & 0.72 \\
ordinal & 3 & original & 0.86 & 0.81 & 0.78 \\
continuous & 3 & original & 0.85 & 0.79 & 0.76 \\
\bottomrule
\end{tabular}
\end{table*}

\paragraph{Design Decisions}
While the slightly improved correlation of the three-point rubric was attractive, especially in the ordinal original-language condition, we ultimately decided to retain the five-point scheme. Conversations with downstream users suggested that the finer-grained rubric offered better interpretability and allowed more flexibility when tuning thresholds for inclusion or exclusion. We likewise opted to keep the source text in its original language: this reduced inference latency and eliminated an entire class of potential failure modes stemming from machine translation errors. Finally, we found that prompting the model for direct ordinal predictions yielded better performance than continuous scoring.

\paragraph{Data Collection}
With the LLM annotation scheme finalized, we used GPT-4o to generate silver annotations for \num{50000} sources, \num{1000} per language.  Each source's text was truncated to \num{2000} characters in the prompt. The annotations were collected via the Azure OpenAI API, with a retry mechanism in place to recover from network errors, as well as malformed or incomplete outputs. In approximately 50 cases, the model failed repeatedly due to Azure content filters; these were annotated by hand to preserve dataset balance across languages.

\subsubsection{Model Selection}

Following silver annotation, we evaluated a variety of models on predicting source quality labels from source text.  We aimed to identify a model that could generalize across languages and quality labels while prioritizing performance, especially precision, on classes 4 and 5.  This prioritization reflects our goal of enabling users to filter to a subset of sources with high-quality text.

\begin{figure}[ht]
   \centering
    \includegraphics[width=0.45\textwidth]{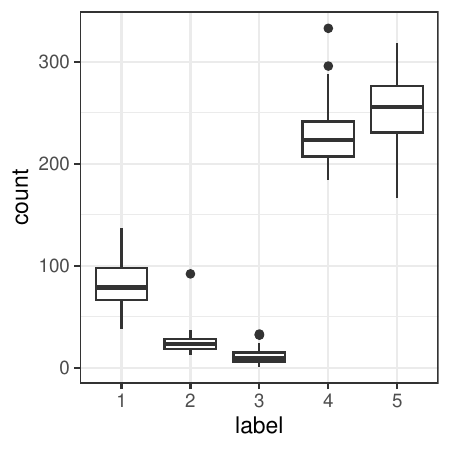}
    \caption{Distribution of each source quality label across languages in the silver training set.}
    \label{fig:silver-train-labels}
\end{figure}

\paragraph{Data Preprocessing}
We split the data into 60/20/20 train/validation/test splits, stratifying with a primary objective of class balance and secondary objective of language balance.
The distribution of silver labels over languages in the training set is depicted in Figure~\ref{fig:silver-train-labels}; there are many more sources in classes 1, 4, and 5 than in classes 2 and 3.

We observed in initial experiments that performance was hindered by the class imbalance, so we upsampled examples from low-frequency classes in the training data.

In our final evaluation of models on the test set, we filter out language-class pairs with less than five instances to reduce estimation variance.
Because of the low frequency of class 3, this filtering results in class 3 not being represented in most languages in our evaluation data.
Languages \emph{without} sufficient support in the test set:
\begin{itemize}
\item \textbf{Class 2}: et 
fa 
id 
km 
mr 
ne 
ps 
vi 
\item \textbf{Class 3}: af 
ar 
cs 
de 
en 
es 
et 
fa 
fi 
fr 
ga 
gl 
he 
hr 
ja 
kk 
ko 
lt 
lv 
mk 
mn 
mr 
my 
ne 
nl 
pl 
ro 
ru 
si 
sl 
sv 
th 
tr 
uk 
vi 
xh 
\end{itemize}

\paragraph{Classical ML Classifiers} 
We first evaluated traditional machine learning classifiers such as logistic regression, random forests, and histogram gradient boosting (HGB). These models were trained on a variety of input features derived from the source texts. Feature types included bag-of-words representations and learned embeddings. For the latter, we experimented with both off-the-shelf and fine-tuned multilingual embedding models. From these, we extracted document-level representations via multiple strategies, including taking the \texttt{[CLS]} token embedding, mean-pooling over token embeddings, and more advanced techniques like attention and learned projections. Among these, simple mean-pooling over embeddings from fine-tuned multilingual embedding models consistently yielded the best results, outperforming both \texttt{[CLS]}-based and learned pooling schemes.  HGB performed best, offering strong baselines across nearly all classes.

\paragraph{Classical ML Regression}
Then, considering the nature of the quality prediction task, we evaluated classical regression models including linear, ridge, Lasso, support vector, random forest, and gradient boosting regression. However, none of these models were able to match the performance of the classifier baselines. Regression predictions often collapsed toward the mean and failed to separate boundary cases.

In addition to standard (continuous) regression models, we evaluated logit and probit ordinal regression models. While this approach was attractive because it allows for fine-tuning threshold boundaries between ordinal classes, the model was limited by its underlying linear structure, and performance lagged behind both traditional models and deep learning methods.

\paragraph{Transformer Classifiers}
We also evaluated transformer-based classifiers, specifically XLM-RoBERTa~\cite{conneau-etal-2020-unsupervised} and Multilingual E5~\cite{wang2024e5}. We used the base model size of XLM-RoBERTa ($\sim$300M parameters) and two e5 model sizes, e5-small ($\sim$100M parameters) and e5-base ($\sim$300M parameters; initialized from XLM-RoBERTa base).\footnote{
    In experiments not reported here, we also tested the large XLM-RoBERTa model, but we found it did not improve performance.
}
Each was equipped with a classification head and fine-tuned directly on our silver-labeled dataset.
These transformer models substantially outperformed the classical ML baselines, particularly in distinguishing mid-range quality labels (2 and 3) and preserving precision at the extremes.

\begin{table*}[ht]
\centering
\begin{tabular}{lrrrrrr}
\toprule
\textbf{Model} & \textbf{Class 1} & \textbf{Class 2} & \textbf{Class 3} & \textbf{Class 4} & \textbf{Class 5} & \textbf{Class 4|5} \\
\midrule
e5-small                             & 0.90 & 0.51 & 0.74 & 0.68 & 0.80 & 0.97 \\
e5-base                              & 0.89 & 0.50 & 0.71 & 0.71 & 0.79 & 0.97 \\
xlmr-base                            & 0.90 & 0.51 & 0.68 & 0.70 & 0.80 & 0.97 \\
\textsuperscript{*}xlmr-base regress & 0.89 & 0.50 & 0.62 & 0.70 & 0.80 & 0.97 \\
\bottomrule
\end{tabular}
\caption{Average (macro) F1 scores (across languages with at least five test examples) of selected models on the silver test set.  4|5 is the pseudo-class consisting of the union of classes 4 and 5.  \textsuperscript{*}selected source quality model.}
\label{tab:source-quality-results}
\end{table*}

\paragraph{Transformer Regression}
Finally, we constructed a transformer-based continuous regression model and (subsequently) fitted ordinal scale thresholds to its outputs.  This approach achieved similar performance to the transformer classifiers while also providing continuous quality score outputs from the initial regression model.  Specifically, we trained XLM-RoBERTa and e5 models as traditional regression models, scaling the training set labels linearly to $[0.1, 0.9]$, and then fit ordinal class thresholds on a subset of the training set.\footnote{
    We initially tried unscaled ($[0, 5]$) labels and labels scaled to $[0, 1]$, but the compressed $[0.1, 0.9]$ scale resulted in the best performance.
}

\paragraph{Evaluation}
We provide transformer model F1 scores on the silver test set, averaged (macro-averaged) across languages with at least five test examples, in Table~\ref{tab:source-quality-results}.
These transformer models achieve similar average F1 overall; the differences are only pronounced in the low-frequency class 3, in which the e5 models perform best and the regression model performs worst (a drop of about 10 points).
The high performance of the regression model in classes 4 and 5, including high precision in those classes (as detailed in Appendix~\ref{appendix:source-quality-statistics}), coupled with the provision of continuous quality scores in parallel alongside ordinal labels, makes the regression model the best fit for source quality labeling.

\subsection{Translation Quality}
\label{sec:translation-analysis}

\megawika~2 uses a newer, larger translation model than \megawika~1, and we expect higher-quality translations.  To demonstrate the improvement, however, we are restricted to reference-free metrics, as we do not have access to ground-truth translations for either \megawika collection.  In this section, we evaluate the fluency of translated passages from each dataset using perplexity as a proxy for fluency.  Specifically, we 
compare the perplexity per token---which we abbreviate as \emph{perplexity}---of randomly sampled English passage translations between datasets and translation models.

\paragraph{Translation Models}
Three different translation models were used between \megawika~1 and \megawika~2:
\begin{enumerate}
    \item \megawika~1 was originally translated using M2M100, which has 418 million parameters~\cite{fan2021beyond}.
    \item \megawika~1 was later translated again using Google Translate.  The version of \megawika~1 described in~\cite{barham2023megawika} and released on HuggingFace (\mwonehf) contains M2M100-translated passages for the 40 largest languages (by article count) and Google Translate translations for the 10 smallest languages, but we have access to M2M100 and Google Translate translations for the full collection.
    \item \megawika~2 is translated using NLLB-200 600M in 8-bit distillation~\cite{nllbteam2022languageleftbehindscaling}. This model is larger and newer than M2M100, and running it is much more scalable than calling the Google Translate API given sufficient GPU resources.
\end{enumerate}

\paragraph{Passage Sampling}
We filter article chunks to those from the first third of each Wikipedia dump.
Articles appearing earlier in a dump are generally older;
informally, we observe that they have often seen more edits and are longer and higher quality.
We also filter out articles with colons in the title, which often indicate special page types like templates (in English, prefixed by \texttt{Template:}) and article discussion pages (in English, prefixed by \texttt{Talk:}).

We then sample 500 target passages (translations) from those articles under the respective translation models (M2M or Google Translate for \megawika~1, NLLB for \megawika~2).  Perplexity under a given language model is negatively correlated with sequence length, regardless of linguistic properties \cite{miaschi-etal-2021-makes}, so to make a fair comparison of translation models, we should only compare passages with similar lengths.
To ensure similar passage lengths, we draw a weighted sample of passages in which we pick a target length $L_{\text{target}}$ and passages with length close to that target have higher weight.  Specifically, if $p$ is a passage and $\ell_p$ is the number of characters in that passage, we sample passages with probability proportional to
\[
\exp\left(-\frac{| \ell_p - L_{\text{target}} |}{L_{\text{target}}}\right)
\]
This distribution has a mode at $\ell_p = L_{\text{target}}$ and falls off sharply in both directions, ensuring the sampled passages are close to the target length on average.  We use a target length of $L_{\text{target}} = 150$.  Implementation details are provided in Appendix~\ref{appendix:translation-extra}.

As seen in Table~\ref{tab:length_stats}, the inner quartiles and means of the sampled target passage lengths from each translation model---locations describing the middle of each distribution---are generally within about 20\% of each other.  M2M target passages are longest on average, followed by NLLB, so samples from these models are likely to have slightly lower perplexity under the same language model.

\begin{table}[ht]
    \centering
    \begin{tabular}{lrrrr}
        \toprule
        Model & Q1 & Median & Mean & Q3 \\
        \midrule
        Google & 81  & 127  & 141  & 179 \\
        M2M  & 117 & 150  & 160  & 193 \\
        NLLB & 65  & 122  & 152  & 204 \\
        \bottomrule
    \end{tabular}
    \caption{Summary statistics of sampled target passage lengths from the three translation models.  \emph{Q1, Q3}: first and third quartiles.}
    \label{tab:length_stats}
\end{table}

\begin{figure*}[ht]
   \centering
    \includegraphics[width=1.0\textwidth]{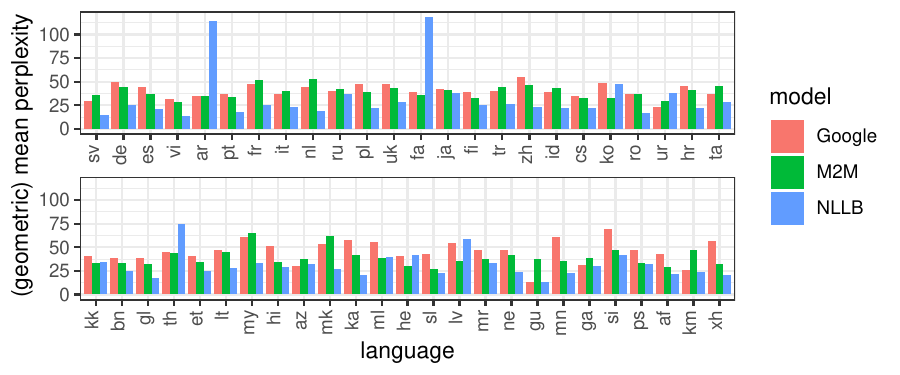}
    \caption{Geometric means of the perplexities for sampled passage translations from each model and language.}
    \label{fig:translation-perplexity}
\end{figure*}

\paragraph{Results}
Figure~\ref{fig:translation-perplexity} shows the geometric mean (the exponentiated negative mean log-likelihood) of the sampled passage translation perplexities for each model and language.\footnote{
    We take the mean in log space, following the definition of perplexity as the geometric mean of token log-likelihoods.  Also note that the mean log-likelihood---that is, negative log perplexity---distributions approximate bell curves, as illustrated in Appendix~\ref{appendix:translation-extra}, specifically Figure~\ref{fig:translation-loglikelihood-density-1} and its companion figures.
}
Note that the perplexity of each \emph{passage} is weighted equally; weighting each token equally (computing corpus-level perplexity) would exacerbate the bias toward longer passages (which have lower perplexity on average).
The NLLB translations used in \megawika~2 have lower average perplexity in most languages than the M2M and Google Translate translations used in \megawika~1.  Notable exceptions are Arabic (ar), Farsi (fa), and to a lesser extent Thai (th), for which NLLB exhibits much higher perplexity than the other two models.

\begin{figure}[ht]
   \centering
    \includegraphics[width=0.5\textwidth]{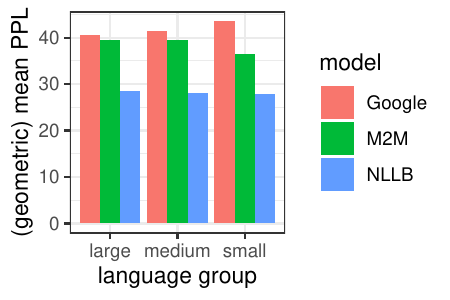}
    \caption{Geometric means of the perplexities (PPL) for sampled passage translations for each model and language group.  Languages are grouped by number of articles in \megawika~1.}
    \label{fig:translation-perplexity-by-language-group}
\end{figure}

In Figure~\ref{fig:translation-perplexity-by-language-group}, we bin the results by language size---the largest 15 languages, middle 20 languages, and 15 smallest languages, by number of articles in \megawika~1---and take the geometric mean within each bin.  This aggregation shows NLLB achieving lowest perplexity overall.  Counterintuitively, it also shows M2M and NLLB achieving lower perplexity for smaller languages---we would expect higher perplexity, taking small Wikipedia projects as a proxy indicator for low-resource languages---while Google Translate achieves higher perplexity for smaller languages as expected.  This finding may point to the limitations of perplexity as a standalone translation evaluation metric.

Histograms of the sampled passage translation log-likelihoods for each language and model are provided in Appendix~\ref{appendix:translation-extra}.

\paragraph{Limitations}
Without reference translations, we only measure translation \emph{fluency}, not \emph{adequacy}~\cite{linguistic2005linguistic}.  However, fluency and accuracy (adequacy) have been shown to be positively correlated at the corpus level (but not for individual segments or sentences)~\cite{lim2024simpson}, and we compare translations' fluency at the corpus level (we compare the average perplexities of Wikipedia corpora).
\section{Related Work}
\label{sec:related-work}

\paragraph{Wikipedia Across AI}
Wikipedia forms the foundation of many datasets for knowledge-intensive NLP tasks~\cite{petroni-etal-2021-kilt}, especially fact checking~\cite{trokhymovych2021wikicheck, chernyavskiy2021whatthewikifact, sathe2021fact, glockner-etal-2024-ambifc} and open-domain question answering (QA)~\cite{singh2013automatic,yang2015wikiqa,chen2017reading,du2018harvesting,lewis-2021}.

Wikipedia is highly multilingual, enabling research on multilingual variants of these tasks~\cite{shivansh-etal-2023-cross, longpre2021mkqa}. Some QA datasets, like XQuAD and MKQA, use automated or human translations of the English versions to derive multilingual queries~\cite{liu-2019,artetxe2019xquad}, while others have used Wikipedia's own cross-lingual article links to unify QA pairs across multilingual contexts~\cite{lewis-2020-mlqa}.  Still others have applied question generation in parallel to Wikipedias in multiple languages, yielding QA collections with disjoint sets of questions across languages, like \tydi{}~\cite{clark2020tydi}.

In addition to multilingual QA, Wikipedia has been used for large-scale multilingual information retrieval.  Much of the research so far extends existing English QA datasets to this setting; for example, Xor-\tydi{} extends \tydi{} to cross-lingual open-retrieval QA on seven languages~\cite{asai2020xor}; 
Mr.~{\abr{t}{\small y}\abr{d}{\small i}} extends \tydi{} to eleven languages~\cite{zhang-etal-2021-mr}, and MiRACL further extends Mr.~\tydi{} to eighteen languages~\cite{miracl, mircal2023}. Because \megawika~2 maps original and translated Wikipedia sentences to scraped sources often written in the original article language, we expect it to support the development of larger and more natural XLQA and XLIR datasets.

Wikipedia can also serve as a source of ground-truth summaries (articles) and corresponding background knowledge to summarize (citations to external sources). \megawika~1 was motivated by this observation, as was WikiSum, a multi-document summarization dataset created by scraping web sources of English Wikipedia articles and using the articles themselves as summaries~\cite{liu2018generating}.  More recently, attention has turned toward verifying individual claims from Wikipedia based on source evidence~\cite{petroni2022improving,kamoi2023wice} and verifying facts generated by search engines by using Wikipedia articles themselves as evidence~\cite{Liu2023EvaluatingVI}.

Wikipedia tables are a rich source of structured data. \citet{bhagavatula2013methods} first presented methods for extracting tables from Wikipedia articles; subsequently, Wikipedia tables have formed the backbone of datasets for semantic parsing, question answering (WikiSQL~\cite{zhong2017seq2sql}, Open-WikiTable~\cite{kweon2023open}), fact verification (TabFact~\cite{chen2019tabfact}), and table-to-text generation (LogicNLG~\cite{chen2020logical}, WikiTableT~\cite{chen2020wikitablet}, and ToTTo~\cite{parikh-etal-2020-totto}).\footnote{
    Most such datasets are derived from HTML renders of Wikipedia articles provided fetched from Wikipedia itself, not generated offline from the raw MediaWiki available in Wikipedia dumps.
}

Infoboxes are also regularly used as a structured data source.  As fixed-format tables at the top-right of Wikipedia articles that capture key facts about the subject (for example, a person’s birth date or occupation), they provide human-curated summaries that are of great value to a number of NLP sub-domains.  For example, WikiBio is a dataset of over \num{700000} Wikipedia biography article texts paired with infoboxes~\cite{lebret2016wikibio}; WikiReading is a machine reading comprehension dataset spanning 18 million infobox-based instances~\cite{hewlett2016wikireading}; and TempTabQA is a temporal QA dataset for answering questions about the chronological events~\cite{gupta2023temptabqa}.

Wikipedia is an interesting data source not just because it is multilingual; it also evolves over time due to the near-constant labor of Wikipedia editors.  Some research thrusts have leveraged this dynamic nature to model changes in knowledge over time.  For example, GrowOVER and EvolvingQA use changes in article text to construct QA benchmarks that evolve over time~\cite{ko-etal-2024-growover,kim-etal-2024-carpe}.
DynaQuest similarly uses changes in infobox data to construct an evolving QA dataset~\cite{lin-etal-2025-dynaquest}.

Finally, Wikipedia is not just used as a read-only source of data; it is also treated as a topic of scientific study or a resource to be improved.  \citet{petroni2022improving} approaches verifiability of claims in Wikipedia as a topic of study and challenge to be addressed, developing a system to recommend alternative sources for Wikipedia article claims.  Similarly, recommender systems have been proposed for adding entity links from one article to another~\cite{gerlach2021multilingual,feith-etal-2024-entity} or proposing articles editors might be interested in contributing to~\cite{moskalenko2020scalable,houtti2024leveraging}.

\paragraph{Wikipedia in Language Modeling}
Wikipedia is also a good source of curated natural language spanning the whole range of human knowledge. As such, Wikipedia formed the basis of some of the earliest medium-scale language model training corpora, like \wikitext-103, which spans \~100 million tokens~\cite{merity2016pointer}. Such corpora focused primarily on English Wikipedia; and they have since been succeeded by large multilingual Wikipedia corpora---like Wiki-40B~\cite{guo2020wiki}---which consist of billions or even tens of billions of tokens.

In the current era of trillion-token pretraining datasets, Wikipedia still forms a high-quality, curated backbone to the natural language portion of all known LLM pretraining curricula~\cite{brown2020language,rae2021scaling,hoffmann2022training,chowdhery2023palm,zhang2022opt,touvron2023llama,achiam2023gpt,touvron2023llama, almazrouei2023falcon,groeneveld2024olmo}. Some recently described pretraining regimes leverage Wikipedia in a kind of low--learning-rate final curriculum stage, where it serves to even out validation loss across corpus subsets~\cite{groeneveld2024olmo,ai22024olmo}.

\paragraph{Related Wikipedia Datasets}

In Table~\ref{table:related-work}, we list the sizes and features of prominent Wikipedia-based NLP datasets most structurally related to \megawika~2.

\begin{table*}[ht]
\centering
\begin{tabular}{p{12em} r r r c c r} \toprule
\textbf{Dataset} & \textbf{Year} & \textbf{Articles} & \textbf{Sources} & \textbf{Open Acc} & \textbf{ML} & \textbf{Storage (GB)} \\ \midrule
    WikiSum \cite{liu2018wikisum}          & 2018 & 2.3~M  & 87~M  & -          & -          & 300  \\
    WikiCatSum \cite{perez2019wikicatsum}  & 2019 & 0.17~M & 24~M  & \Checkmark & -          & 4.8  \\
    Hiersum \cite{liu2019hiersum}          & 2019 & 1.7~M  & -     & - & -          & 6.9  \\
    WebBrain-Raw \cite{qian2023webbrain}   & 2023 & 15~M   & 260~M & \Checkmark\textsuperscript{*} & -          & 2800 \\ \midrule
    \megawika~1 \cite{barham2023megawika}  & 2023 & 13~M   & 71~M  & \Checkmark & \Checkmark & 1000  \\
    \ -- English subset & & 0.84~M & 3.2~M\textsuperscript{\textdagger} & & & 56 \\
\midrule
    \textbf{\megawika~2}                            & 2025 & 77~M   & 172~M & \Checkmark & \Checkmark & 2200 \\
    \ -- English subset & & 9.8~M & 57~M & & & 580 \\ \bottomrule
\end{tabular}
\caption{Sizes of Wikipedia-based NLP corpora ordered by publication date.  \textbf{ML}: multilingual; \textbf{Open Acc}: open access data.  M means million.  Sizes were computed \emph{uncompressed} copies of the data where possible.  \textsuperscript{*}\href{https://github.com/qhjqhj00/WebBrain-Data/blob/ae4620d7fbffc938dbb9182cb61bc9409b351ad1/README.md\#application-form}{The WebBrain GitHub page} indicates researchers may apply for access to the full dataset, but we applied in early 2025 and have not received a response at the time of writing.  \textsuperscript{\textdagger}All web citations in the \megawika~1 English subset on HuggingFace have non-empty text extracted, so the true web citation count may be much higher.}
\label{table:related-work}
\end{table*}

WebBrain~\cite{qian2023webbrain} was developed more or less concurrently with \megawika~1, and it is the dataset most comparable to \megawika~2 in size.  WebBrain focused on article generation in an open-domain setting. They also built a novel dataset, but like~\cite{lewis-2021} use English Wikipedia only.

In contrast to \megawika~2, WebBrain:
\begin{itemize}
    \item collects only English articles;
    \item collects references at the article level, attaching them not to individual words or sentences, but to the introduction section of the article;
    \item preprocesses the scraped references differently, filtering out non-English references, references deemed to be incorrectly cited, etc.; and
    \item only contains selected passages from scraped references, namely those with the highest $P_{ST}$, a metric of relatedness~\cite{piktus2021oyster}.
\end{itemize}

WebBrain-Raw consists of 15 million English articles containing 260 million references (web citations) while \megawika~2 has 10 million English articles containing 57 million web citations. Thus, WebBrain-Raw has many more references per article than \megawika~2. While we could not find the full WebBrain-Raw data set, a 500-article sample was provided in the supplementary materials of the WebBrain paper.  We looked up article titles from this sample in \megawika~2 and found that, on this sample, \megawika~2 has slightly more web citations per article than WebBrain-Raw has references per article (22 web citations compared to 18 references).  Thus, we would expect the corpus statistics to be similar.  The WebBrain-Raw sample does appear consistent with a uniform-at-random sample; it has an average of 18 references per article, compared to 17 references per article in the entire WebBrain-Raw data set.  Details of this analysis are provided in Appendix~\ref{appendix:webbrain-sample-analysis}.

Because we filter out some redirects, category pages, and stub pages---particularly in English---we expect to have less than the full number of articles. Inspecting the May 1, 2024 Wikipedia dump for English, we find there are 24 million articles in total.\footnote{
    Article IDs appear to be auto-incrementing, and the first ID in the dump is \num{10} while the last is \num{76788691}, suggesting there have been over 76 million articles in English Wikipedia over its recorded lifespan.  Dumps do not contain entries for deleted articles.
}\footnote{
    In the May 1, 2024 Wikipedia XML dump, about \num{400000} articles, or 4\%, were created after April 10, 2023, the date of publication of WebBrain, so we would expect a 4\% increase in articles, not accounting for deletions.  In \megawika~2 English, articles later in the dump (newer articles, approximately) tend to have slightly fewer citations (Spearman rank correlation $-0.05$), so we do not expect a proportional increase in citations.
}
However, we expect the average citation count of the filtered-out pages to be close to zero, as they are generally incomplete or irregular articles.  The drop in citation count from WebBrain-Raw to \megawika~2 is concerning, but the higher citation count in \megawika~2 suggests a comparison of \megawika~2 processing on all English articles with the full WebBrain-Raw dataset may be needed to uncover the cause of the discrepancy.

\section{Future Work}
\label{sec:future-work}

\subsection{\wikitext Parsing Improvements}

\megawika~2 collects the raw \wikitext of structured article elements like tables, infoboxes, and multimedia blocks. Tables and infoboxes in particular represent dense, curated troves of structured knowledge, offering valuable signals for tasks like fact verification, question answering, table-to-text generation. However, in their unparsed form, they are difficult to use at best. Parsing these elements is complicated by Wikipedia’s powerful but idiosyncratic template system, which supports arbitrarily recursive template nesting, conditional logic (including loops and other iterative control structures), and embedded Lua scripting. In future work, we aim to evaluate the feasibility of using the official MediaWiki parser for full template resolution.  We will also explore more lightweight alternatives that can still extract meaningful structure---such as infobox field-value pairs or table schemas---without requiring full template expansion.

Wikipedia's template language also drives the structure and semantics of inline citations. As a result, improving citation extraction accuracy remains a challenge in some languages, particularly for articles that rely on families of lesser-known or language-specific citation templates. While \megawika~2 expanded coverage beyond basic `<ref>` tags to include many citation templates (for example, `{{cite}}`, `{{citation}}`, `{{rp}}`), our own analyses have identified additional high-impact families (for example, `{{harv}}`, `{{sfn}}`, `{{webbref}}`, `{{bookref}}`, etc.) that our current approach misses. Moreover, citation practices vary \textit{significantly} across languages. In future work, we plan to expand the citation template whitelist, evaluate multilingual citation recall, and improve extraction of source snippets---particularly those embedded in `quote` parameters or named references. These changes are critical to improving citation coverage and consistency across the full corpus.

\subsection{Source Scraping Improvements}

Improving the source download success rate may be possible by using a proxy pool or otherwise taking care to avoid rate limiting.

Source extraction could be improved in many ways.  Source extraction currently filters out pages with few whitespace-separated ``words,'' often irrelevant page content like 404 ``page not found'' content, paywall messages, and minimized JavaScript for dynamically generated content.  This filter should be adjusted to better accommodate languages like Chinese that do not generally use separate whitespace characters to delimit words and sentences.
Separately, it may be possible to use a headless browser to recover dynamically generated source content.

Source extraction could also be extended to other content types, most notably PDF files.  While PDF parsing is significantly more computationally expensive, a preliminary analysis suggests that sources are about 95\% HTML web pages by size, so the total workload may be manageable.

\subsection{Further Enrichments}

Many kinds of enrichments that could be added to reduce the preprocessing needed by downstream users.  A categorization of article types, even a binary categorization of regular articles and ``special'' articles like talk pages, template pages, and others, would help most users filter to articles with typical content and structure.  We currently use a heuristic, filtering out articles with a colon or forward slash in the title, but the precision and likely the recall of this filter could be improved.

Since Wikipedia is often treated as a collection of natural-language articles written by humans, it could also be useful to identify which articles were originally written (by translating articles in other Wikipedia projects or otherwise) by a bot.  However, the increasing ubiquity and accessibility of large language models in day-to-day life complicates this categorization, as users may be submitting AI-authored articles and edits.

Finally, we retrieve cross-lingual links and creation dates for articles from the Wikipedia Action API, but these links do not follow redirects, and interpreting article metadata from the Action API in the presence of a current or historical redirect is fraught.  It would be helpful to provide more clarity about the limitations of this metadata, and perhaps to provide redirect-resolved metadata alongside it.  Details about which articles are redirects, and what titles redirect to an article, could also be useful enrichments.

\subsection{Incremental Updates}

We have downloaded Wikipedia dumps from the first of each month, from May 2024 to March 2025 (the time of writing), for all fifty \megawika languages.  The exceptions to date are January 1, 2025 and July 1, 2025, for which we have no English dumps due to upstream Wikipedia dump process failures.  In its place, we have downloaded the December 20, 2024, January 23, 2025, June 20, 2025, and July 20, 2025 dumps for English.  We plan to continue downloading dumps from the first of each month whenever possible.

Due to time constraints, we have only run the extraction, translation, and enrichment pipelines on the May 2024 dumps at the time of writing.  To facilitate processing subsequent dumps, we plan to further improve the efficiency of the delta mode by caching key fields of each \megawika~2 article.\footnote{
    Specifically, we intend to cache citations (including source text and text quality), translations, creation dates.  Source content and article creation dates are produced by accessing the source web pages and the Wikipedia Action API, respectively.  These web resources may change over time; source web pages may be updated or removed, and (we suspect) creation dates may change when an article is converted from a regular article to a redirect or vice versa.  However, for the sake of scope, we assume these resources/fields do not change unless their explicit input from the pipeline changes.
}
This improvement will allow us to avoid re-running the most resource-intensive functions, dramatically reducing the time and computational resources needed to run the delta pipelines.  We plan to retain the existing article-level delta mode logic, as it dramatically reduces the storage space required to run the delta pipelines.

\section{Conclusions}
\label{sec:conclusion}

We have introduced \megawika~2, a large, multilingual dataset of structured Wikipedia articles containing rich elements, scraped sources for web citations, English translations of article text, and further enrichments supporting analyses across language and time.  \megawika~2 improves over the original \megawika qualitatively and quantitatively, and it focuses on tasks in the neighborhood of article generation instead of question answering.  The \megawika~2 data processing pipeline is designed to be run incrementally as new Wikipedia dumps are released, and in future work we plan to augment the dataset with monthly updates alongside, not overwriting, the original data, facilitating analyses of data drift in LLMs~\cite{fleshman2024adapterswap} and other temporal studies.

\section{Limitations}
\label{sec:limitations}

Although our dataset covers many languages beyond English, there are many more languages represented in Wikipedia that we are not ingested.  Inclusion of additional languages in \megawika~2 would be valuable.

As mentioned in the parsing subsection of Section~\ref{sec:data-processing}, editing conventions vary across Wikipedia projects (languages).  Wikipedia readers' use cases---their reasons for using Wikipedia---also vary across languages~\cite{lemmerich2019why}.  This variation problematizes \megawika~2, which (like \megawika~1) is a an effort to collect Wikipedia articles and their sources and postprocess them into a standard format.  Normalizing the data in this way may obscure important information, especially for sociolinguistic research or development of personalized editing tools.  Similarly, presenting Wikipedia and its sources as a unified dataset encourages research in the vein of automation and consolidation and contributes to its high valuation in academia and industry; by doing so, we incur an opportunity cost of \emph{not} enabling research in other, perhaps opposing or orthogonal directions.

Due to the scale and volunteer-written nature of Wikipedia, article quality varies, and this random variation/error is compounded by the complexity of source scraping.  Some source text may contain information that is incorrect, irrelevant, or problematically formatted.  Despite these limitations in the underlying Wikipedia data, \citet{barham2023megawika} show that it is usable a number of NLP tasks.  We expect the data processing quality enhancements introduced in \megawika~2 and addition of source text quality estimates to further improve utility.


\section{Acknowledgments}
\label{sec:acknowledgments}

\megawika~2 was based on the extensive work presented in \citet{barham2023megawika} and the contributions of many other people.  Alexander Martin has also contributed substantially to the \megawika~2 data processing pipeline, and Alexander, Orion Weller, Kathryn Ricci, Zhengping Jiang, Miriam Wanner, and William Walden have provided valuable feedback on initial versions of the data.  Thank you to Marc Marone for assisting with the FineWeb-Edu analysis, and thank you to Benjamin Van Durme's lab for providing thoughtful feedback on an earlier version of this manuscript.  Data processing and transfer itself would not have been possible without the ongoing support of IDIES and DSAI, especially Gerard Lemson, Lance Joseph, and Dmitry Medvedev.  Finally, we are indebted to the Wikimedia Foundation and the Wikimedia community for creating, operating, and continually improving Wikipedia and its sister sites, a massive constellation of resources making \megawika~2 conceivable.

The opinions expressed in this paper are those of the authors; they do not reflect the stance of the Wikimedia Foundation or any other entity.

\clearpage

\bibliography{main}

\newpage

\appendix

\section{Data Schema}
\label{appendix:schema}



MegaWika 2 is structured as a collection of JSON-lines ``chunk'' files organized by Wikipedia language.  Each chunk file contains a collection of Article objects, one (JSON-encoded) Article per line.  What follows is documentation for each type in the schema, starting with Article.

\subsection{Article}
\label{appendix:schema-article}

\begin{description}
\item[title] (string): Article title
  \begin{itemize}
  \item {\RaggedLeft\footnotesize Ex: \texttt{"Les Hauts de Hurlevent est l'unique roman d'Emily Brontë ..."}}
  \end{itemize}
\item[wikicode] (string): Wikimedia source code for article
  \begin{itemize}
  \item {\RaggedLeft\footnotesize Ex: \texttt{"<div id=\textbackslash\{\}"mp\_header\textbackslash\{\}" class=\textbackslash\{\}"mp\_outerbox\textbackslash\{\}"> ..."}}
  \end{itemize}
\item[hash] (string): Hash of title and content
  \begin{itemize}
  \item {\RaggedLeft\footnotesize Ex: \texttt{"2c0c3bfb0493fb8ddd5661..."}}
  \end{itemize}
\item[last\_revision] (string): Datetime of last revision
  \begin{itemize}
  \item {\RaggedLeft\footnotesize Ex: \texttt{"2023-12-03T10:50:40Z"}}
  \end{itemize}
\item[first\_revision] (string | null): Datetime of initial article creation, if it could be retrieved.
  \begin{itemize}
  \item {\RaggedLeft\footnotesize Ex: \texttt{"2023-09-04T08:19:40Z"}}
  \end{itemize}
\item[first\_revision\_access\_date] (string | null): Datetime first revision was retrieved from Wikipedia Action API
  \begin{itemize}
  \item {\RaggedLeft\footnotesize Ex: \texttt{"2023-12-03T10:55:40Z"}}
  \end{itemize}
\item[cross\_lingual\_links] (object[string, string] | null): A dictionary mapping this article onto articles on the same topic in other languages; keys represent language codes, values represent the title of the article in that language.
  \begin{itemize}
  \item {\RaggedLeft\footnotesize Ex: \texttt{\{"en": "Wuthering Heights", "es": "Cumbres Borrascosas"\}}}
  \end{itemize}
\item[cross\_lingual\_links\_access\_date] (string | null): Datetime cross-lingual links were retrieved from Wikipedia Action API
  \begin{itemize}
  \item {\RaggedLeft\footnotesize Ex: \texttt{"2023-12-03T10:56:40Z"}}
  \end{itemize}
\item[text] (string): Natural-language text of article
  \begin{itemize}
  \item {\RaggedLeft\footnotesize Ex: \texttt{"Les Hauts de Hurlevent est l'unique roman d'Emily Brontë ..."}}
  \end{itemize}
\item[elements] (array[\hyperref[appendix:schema-heading]{Heading} | \hyperref[appendix:schema-table]{Table} | \hyperref[appendix:schema-infobox]{Infobox} | \hyperref[appendix:schema-paragraph]{Paragraph} | \hyperref[appendix:schema-math]{Math} | \hyperref[appendix:schema-code]{Code} | \hyperref[appendix:schema-preformatted]{Preformatted}]): Article structure: paragraphs, text and citation elements, etc.
\item[excerpts\_with\_citations] (array[\hyperref[appendix:schema-excerptwithcitations]{ExcerptWithCitations}]): A list of all citations from the article and the associated text excerpts they appear in.  This data is a postprocessed subset of the data in the elements list and is provided for convenience.
\end{description}

\subsection{Citation}
\label{appendix:schema-citation}

\begin{description}
\item[content] (string): Citation content
  \begin{itemize}
  \item {\RaggedLeft\footnotesize Ex: \texttt{"<ref>\{\{Citation |last=Thomas |first=Darcy |year=2013 ..."}}
  \end{itemize}
\item[char\_index] (integer): Character index of this citation in the enclosing sentence or excerpt
  \begin{itemize}
  \item {\RaggedLeft\footnotesize Ex: \texttt{39}}
  \end{itemize}
\item[name] (string | null): Optional citation name
  \begin{itemize}
  \item {\RaggedLeft\footnotesize Ex: \texttt{null}}
  \item {\RaggedLeft\footnotesize Ex: \texttt{"Thomas2013"}}
  \end{itemize}
\item[url] (string | null): Extracted URL, if web citation
  \begin{itemize}
  \item {\RaggedLeft\footnotesize Ex: \texttt{"https://example.com/emily-bronte/..."}}
  \end{itemize}
\item[source\_text] (string | null): Extracted source text, if source download and extraction succeeded
  \begin{itemize}
  \item {\RaggedLeft\footnotesize Ex: \texttt{"Emily Brontë avait deux sœurs ..."}}
  \end{itemize}
\item[source\_code\_content\_type] (string | null): Downloaded source code content type, if download succeeded and content-type header was received
  \begin{itemize}
  \item {\RaggedLeft\footnotesize Ex: \texttt{"text/html"}}
  \item {\RaggedLeft\footnotesize Ex: \texttt{"text/html; charset=ISO-8859-1"}}
  \end{itemize}
\item[source\_code\_num\_bytes] (integer | null): Not used
  \begin{itemize}
  \item {\RaggedLeft\footnotesize Ex: \texttt{null}}
  \end{itemize}
\item[source\_code\_num\_chars] (integer | null): Size of downloaded source code in characters, if source download succeeded and code can be decoded as text
  \begin{itemize}
  \item {\RaggedLeft\footnotesize Ex: \texttt{100000}}
  \end{itemize}
\item[source\_download\_date] (string | null): Datetime source code was downloaded from the web
  \begin{itemize}
  \item {\RaggedLeft\footnotesize Ex: \texttt{"2023-12-03T10:50:40Z"}}
  \end{itemize}
\item[source\_download\_error] (string | null): Source download error message, if there was an error
  \begin{itemize}
  \item {\RaggedLeft\footnotesize Ex: \texttt{null}}
  \item {\RaggedLeft\footnotesize Ex: \texttt{"Download is too large (2.4 MB)"}}
  \item {\RaggedLeft\footnotesize Ex: \texttt{"ConnectTimeoutError: ..."}}
  \end{itemize}
\item[source\_extract\_error] (string | null): Source extraction error message, if there was an error
  \begin{itemize}
  \item {\RaggedLeft\footnotesize Ex: \texttt{null}}
  \item {\RaggedLeft\footnotesize Ex: \texttt{"Text is too short (50 words)"}}
  \item {\RaggedLeft\footnotesize Ex: \texttt{"Exception: ..."}}
  \end{itemize}
\item[source\_snippet] (string | null): A relevant snippet from the source document, excertped manually by Wikipedia editor; stored in the `quote` field of the relevant citation templates.
  \begin{itemize}
  \item {\RaggedLeft\footnotesize Ex: \texttt{"Emily Brontë avait deux sœurs"}}
  \end{itemize}
\item[source\_quality\_label] (integer | null): An integer between 1 and 5 representing the predicted relevance and quality of the text extracted from the source page: 1 is irrelevant content like 404 text and paywalls, 2 is likely irrelevant or unreadable content like a list of headlines or mangled table, 3 is potentially relevant content like a book abstract, 4 is likely relevant content but with some quality issues, and 5 is relevant content that is well-formatted.
  \begin{itemize}
  \item {\RaggedLeft\footnotesize Ex: \texttt{4}}
  \end{itemize}
\item[source\_quality\_raw\_score] (number | null): The raw score output by the source quality regression model, generally between 0 and 1.  The source quality label is computed from the raw score and has a monotonic but non-linear relationship.
  \begin{itemize}
  \item {\RaggedLeft\footnotesize Ex: \texttt{0.8}}
  \end{itemize}
\end{description}

\subsection{CitationNeeded}
\label{appendix:schema-citationneeded}

\begin{description}
\item[type] (const string = \texttt{"citation-needed"}): Used to differentiate from other element types
\item[content] (string): Citation-needed element content
  \begin{itemize}
  \item {\RaggedLeft\footnotesize Ex: \texttt{"\{\{Citation needed|date=September 2015\}\}"}}
  \end{itemize}
\item[char\_index] (integer): Character index of this citation-needed in the enclosing sentence or excerpt
  \begin{itemize}
  \item {\RaggedLeft\footnotesize Ex: \texttt{39}}
  \end{itemize}
\end{description}

\subsection{Code}
\label{appendix:schema-code}

\begin{description}
\item[type] (const string = \texttt{"code"}): Used to differentiate from other element types
\item[language] (string | null): Code language (as used for syntax highlighting)
  \begin{itemize}
  \item {\RaggedLeft\footnotesize Ex: \texttt{"cpp"}}
  \end{itemize}
\item[content] (string): Code block content
  \begin{itemize}
  \item {\RaggedLeft\footnotesize Ex: \texttt{"int main() \{ ..."}}
  \end{itemize}
\end{description}

\subsection{ExcerptWithCitations}
\label{appendix:schema-excerptwithcitations}

\begin{description}
\item[text] (string): The text of three consecutive sentences from an article
  \begin{itemize}
  \item {\RaggedLeft\footnotesize Ex: \texttt{"Les Hauts de Hurlevent est .... défis à la culture victorienne."}}
  \end{itemize}
\item[translated\_text] (string | null): English translation of the excerpt text, if not in English Wikipedia
  \begin{itemize}
  \item {\RaggedLeft\footnotesize Ex: \texttt{"Wuthering Heights is .... challenges to Victorian culture."}}
  \end{itemize}
\item[citations] (array[\hyperref[appendix:schema-citation]{Citation}]): Citation(s) appearing in the final sentence of this excerpt
\end{description}

\subsection{Heading}
\label{appendix:schema-heading}

\begin{description}
\item[type] (const string = \texttt{"heading"}): Used to differentiate from other element types
\item[text] (string): Heading text
  \begin{itemize}
  \item {\RaggedLeft\footnotesize Ex: \texttt{"Personnages"}}
  \end{itemize}
\item[translated\_text] (string | null): English translation of heading text, if not in English Wikipedia
  \begin{itemize}
  \item {\RaggedLeft\footnotesize Ex: \texttt{"Characters"}}
  \end{itemize}
\item[level] (integer): Heading level (1 being top-level/most general, 6 being bottom-level/most specific)
  \begin{itemize}
  \item {\RaggedLeft\footnotesize Ex: \texttt{2}}
  \end{itemize}
\item[citations] (array[\hyperref[appendix:schema-citation]{Citation}]): Citations appearing in this heading
\item[citations\_needed] (array[\hyperref[appendix:schema-citationneeded]{CitationNeeded}]): Citation-needed elements appearing in this heading
\end{description}

\subsection{Infobox}
\label{appendix:schema-infobox}

\begin{description}
\item[type] (const string = \texttt{"infobox"}): Used to differentiate from other element types
\item[content] (string): Infobox content
  \begin{itemize}
  \item {\RaggedLeft\footnotesize Ex: \texttt{"\{\{Infobox Livre\textbackslash\{\}n| auteur = Emily Brontë\textbackslash\{\}n...\textbackslash\{\}n\}"}}
  \end{itemize}
\end{description}

\subsection{Math}
\label{appendix:schema-math}

\begin{description}
\item[type] (const string = \texttt{"math"}): Used to differentiate from other element types
\item[content] (string): Math block content
  \begin{itemize}
  \item {\RaggedLeft\footnotesize Ex: \texttt{"\textbackslash\{\}\textbackslash\{\}sin 2\textbackslash\{\}\textbackslash\{\}pi x + \textbackslash\{\}\textbackslash\{\}ln e ..."}}
  \end{itemize}
\end{description}

\subsection{Paragraph}
\label{appendix:schema-paragraph}

\begin{description}
\item[type] (const string = \texttt{"paragraph"}): Used to differentiate from other element types
\item[sentences] (array[\hyperref[appendix:schema-sentence]{Sentence}]): List of sentences in this paragraph
\end{description}

\subsection{Preformatted}
\label{appendix:schema-preformatted}

\begin{description}
\item[type] (const string = \texttt{"preformatted"}): Used to differentiate from other element types
\item[content] (string): Preformatted block content
  \begin{itemize}
  \item {\RaggedLeft\footnotesize Ex: \texttt{"\_\_\_\_\textbackslash\{\}n|DD|\_\_\_\_T\_\textbackslash\{\}n|\_ |\_\_\_\_\_|<\textbackslash\{\}n  @-@-@-oo\textbackslash\{\}\textbackslash\{\}\textbackslash\{\}n"}}
  \end{itemize}
\end{description}

\subsection{Sentence}
\label{appendix:schema-sentence}

\begin{description}
\item[text] (string): Sentence text content
  \begin{itemize}
  \item {\RaggedLeft\footnotesize Ex: \texttt{"Les Hauts de Hurlevent est l'unique roman d'Emily Brontë."}}
  \end{itemize}
\item[translated\_text] (string | null): English translation of sentence text content, if not in English Wikipedia
  \begin{itemize}
  \item {\RaggedLeft\footnotesize Ex: \texttt{"Wuthering Heights is the only novel by Emily Brontë."}}
  \end{itemize}
\item[trailing\_whitespace] (string): If the sentence was originally followed by whitespace, this will be a space. If the sentence was not followed by whitespace (for example, if it was followed by a quotation mark), this will be the empty string.
  \begin{itemize}
  \item {\RaggedLeft\footnotesize Ex: \texttt{" "}}
  \item {\RaggedLeft\footnotesize Ex: \texttt{""}}
  \end{itemize}
\item[citations] (array[\hyperref[appendix:schema-citation]{Citation}]): Citations appearing in this sentence
\item[citations\_needed] (array[\hyperref[appendix:schema-citationneeded]{CitationNeeded}]): Citation-needed elements appearing in this sentence
\end{description}

\subsection{Table}
\label{appendix:schema-table}

\begin{description}
\item[type] (const string = \texttt{"table"}): Used to differentiate from other element types
\item[content] (string): Table content
  \begin{itemize}
  \item {\RaggedLeft\footnotesize Ex: \texttt{"\{| class=\textbackslash\{\}"wikitable\textbackslash\{\}"\textbackslash\{\}n|+ Personnages\textbackslash\{\}n|-\textbackslash\{\}n! Nom !! ...\textbackslash\{\}n...\textbackslash\{\}n|\}"}}
  \end{itemize}
\end{description}

\section{Additional Citation Analysis}
\label{appendix:citation-analysis-extra}

\begin{figure*}[ht]
   \centering
    \includegraphics[width=1.0\textwidth]{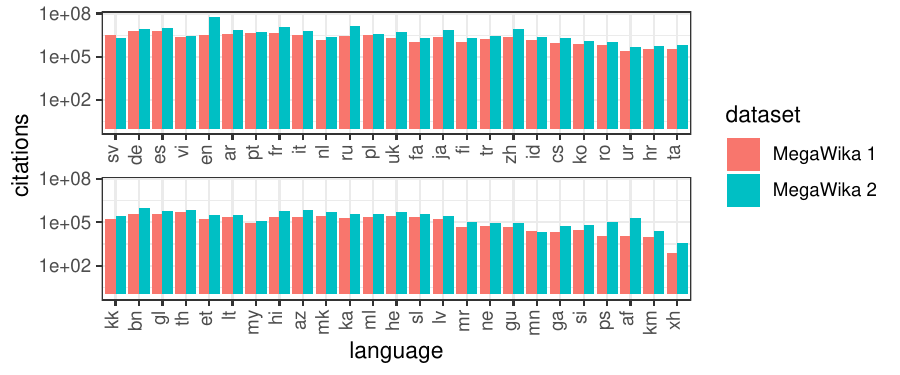}
    \caption{Number of web citations (log scale) per language in \megawika~1 and \megawika~2.}
    \label{fig:web-citations}
\end{figure*}

\begin{figure*}[ht]
   \centering
    \includegraphics[width=1.0\textwidth]{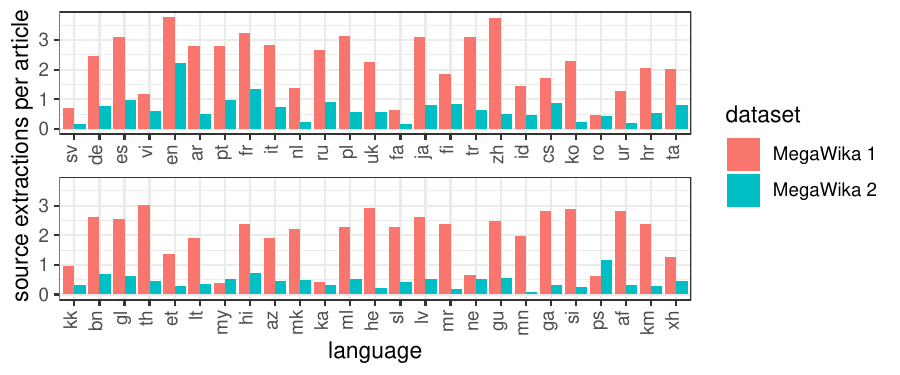}
    \caption{Average number of sources extracted (with non-empty extracted text) \emph{per article} for each language in \megawika~1 and \megawika~2.}
    \label{fig:source-extractions-per-article}
\end{figure*}

\begin{figure*}[ht]
   \centering
    \includegraphics[width=1.0\textwidth]{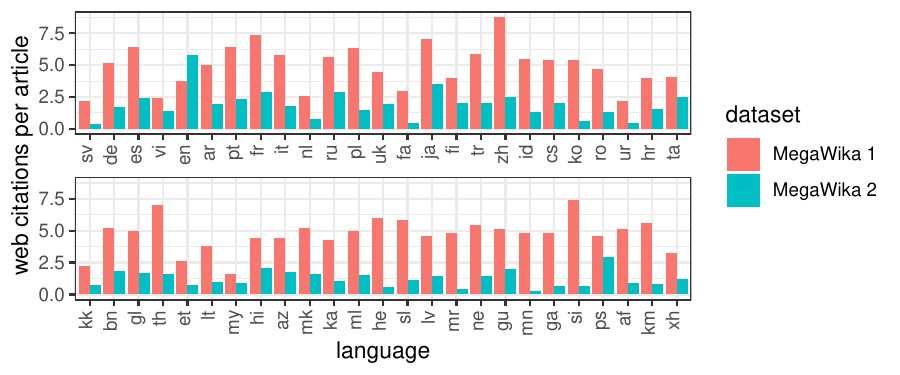}
    \caption{Average number of web citations per article for each language in \megawika~1 and \megawika~2.}
    \label{fig:web-citations-per-article}
\end{figure*}

\begin{figure*}[ht]
   \centering
    \includegraphics[width=1.0\textwidth]{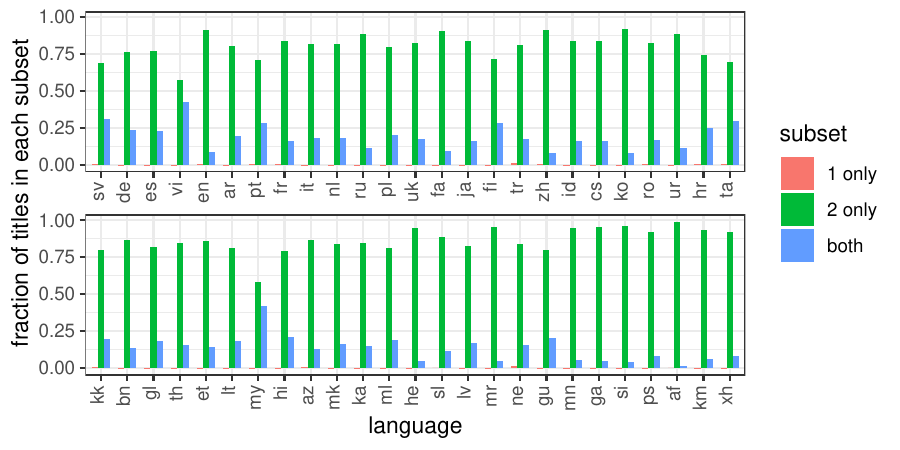}
    \caption{Fraction of article titles in both \megawika~1 and \megawika~2, in \megawika~1 only, or in \megawika~2 only (for each language).}
    \label{fig:mw1-mw2-venn-diagrams}
\end{figure*}

\begin{figure*}[ht]
   \centering
    \includegraphics[width=1.0\textwidth]{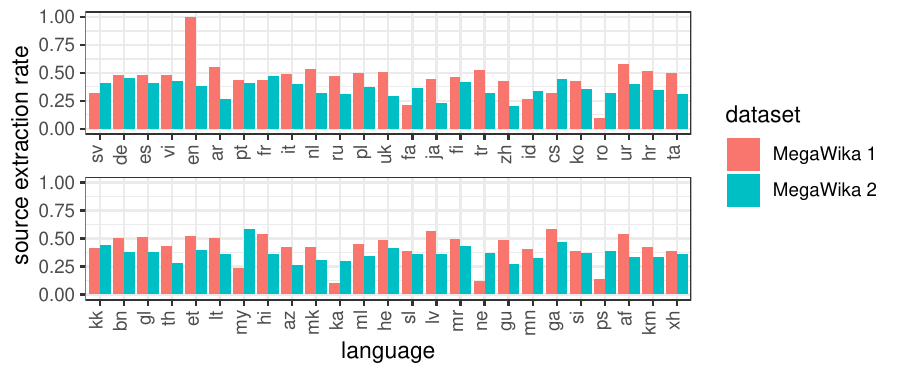}
    \caption{Source extraction rates (fraction of web citation sources with non-empty text extracted) for each language in \megawika~1 and \megawika~2.}
    \label{fig:source-extraction-rates}
\end{figure*}

\begin{figure*}[ht]
   \centering
    \includegraphics[width=1.0\textwidth]{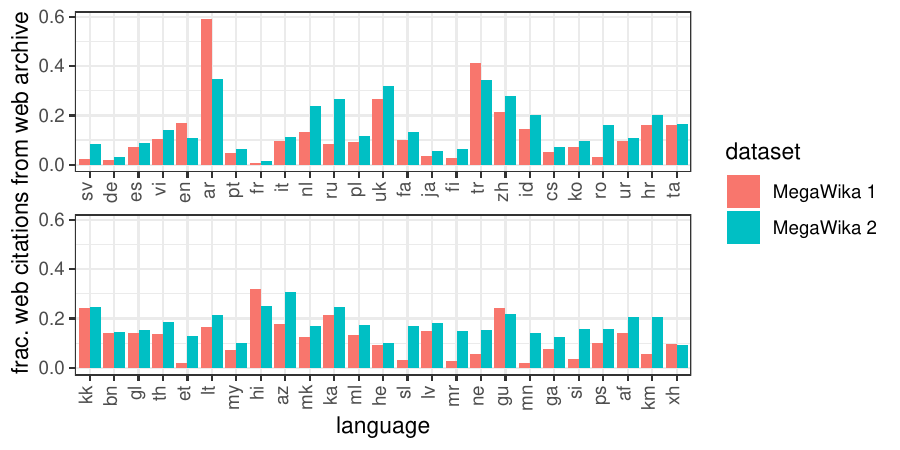}
    \caption{Fraction of web citations that point to sources on web.archive.org in \megawika~1 and \megawika~2.}
    \label{fig:web-archive-web-citations}
\end{figure*}

\begin{figure*}[ht]
   \centering
    \includegraphics[width=1.0\textwidth]{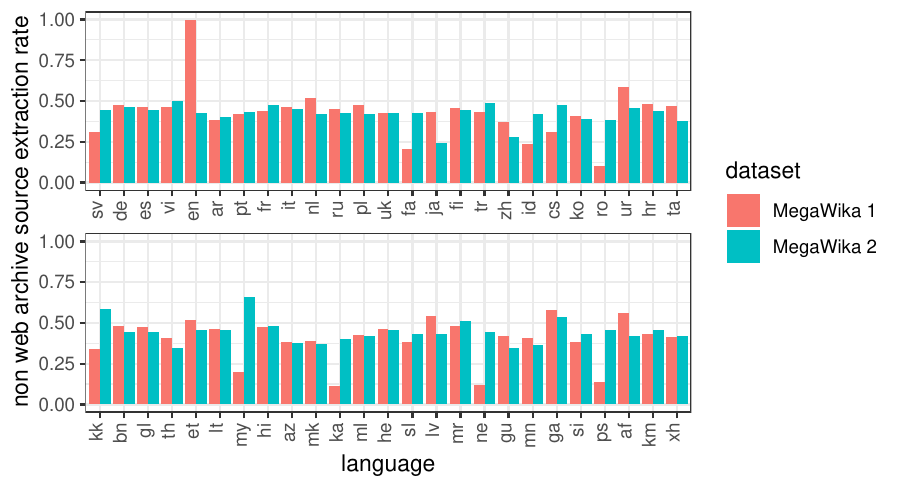}
    \caption{Source extraction rates for non-web.archive.org sources for each language in \megawika~1 and \megawika~2.}
    \label{fig:non-web-archive-source-extraction-rate}
\end{figure*}

\begin{figure*}[ht]
   \centering
    \includegraphics[width=1.0\textwidth]{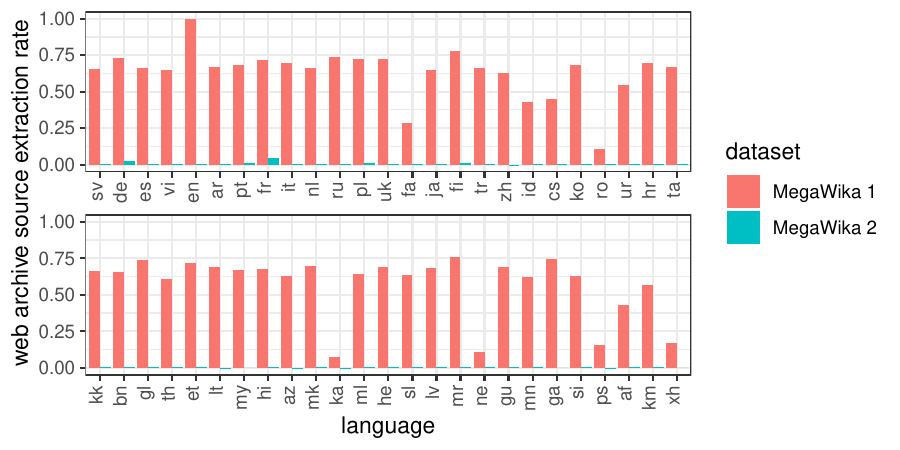}
    \caption{Source extraction rates for web.archive.org sources for each language in \megawika~1 and \megawika~2.}
    \label{fig:web-archive-source-extraction-rate}
\end{figure*}

Additional plots describing differences in citation counts and rates between \megawika~1 and \megawika~2:
\begin{itemize}
\item Web citation counts (Figure~\ref{fig:web-citations};
\item Average sources extracted per article (Figure~\ref{fig:source-extractions-per-article});
\item Average web citations per article (Figure~\ref{fig:web-citations-per-article};
\item Fraction of article titles in \megawika~1 only, in \megawika~2 only, or in \megawika~1 and \megawika~2 (Figure~\ref{fig:mw1-mw2-venn-diagrams});
\item Source extraction rates (Figure~\ref{fig:source-extraction-rates});
\item Fraction of web citations that point to web.archive.org pages (Figure~\ref{fig:web-archive-web-citations});
\item Source extraction rates for non-web.archive.org sources only (Figure~\ref{fig:non-web-archive-source-extraction-rate});
\item Source extraction rates for web.archive.org sources only (Figure~\ref{fig:web-archive-source-extraction-rate}).
\end{itemize}
See Section~\ref{sec:citation-analysis} for main citation analysis.

\section{Source Quality}
\label{appendix:source-quality-extra}

\subsection{Silver Annotation Prompt}
\label{appendix:silver-annotation-prompt}

We used the following prompt to GPT-4o to collect silver source quality annotations:

\lstdefinelanguage{Prompt}{
  basicstyle=\ttfamily\small,
  breaklines=true,
  keepspaces=true,
  columns=flexible,
  breakatwhitespace=true,
  showstringspaces=false,
  escapeinside={(*@}{@*)},
}

\lstset{
  frame=single,
  rulecolor=\color{black},
  backgroundcolor=\color{gray!5},
  framerule=0.5pt,
  xleftmargin=1em,
  xrightmargin=1em,
  aboveskip=1em,
  belowskip=1em,
}

\begin{lstlisting}[language=Prompt, columns=fullflexible]
You're a human analyst who is leveraging
sources to write an article. Here are the
first {N} characters of a scraped source
article you'd *like* to use:
"""
{source_article}
"""
If it helps, the source article above
was scraped from `{url}`.

Before you use the article in your work,
however, you first need to make a
determination as to whether the source
scrape is of high enough quality. We
measure quality on a scale from 1 to 5,
with 1 being the lowest, and 5 being the
highest. Here are descriptions of the
categories:
1: 'Not relevant to indended source material; may be text from a paywall, HTTP 404 or other error message, CAPTCHA page, site navigation menu, promotional content or link farm, etc.',
2: 'Unlikely to be relevant to indended source material and interpretable; may be a list of unrelated article excerpts or a table that was mangled.',
3: 'May or may not represent intended source text; may be a book abstract or website "about" page that suggests the scraper might have been redirected away from the intended content.',
4: 'Represents intended source text, but with some readability or quality issues; may have distracting formatting issues or a significant amount of repeated or irrelevant text.',
5: 'Represents intended source material with little or no readability and quality issues; may include a few markdown-formatted links, but is easy to read and interpret.'

What score should the source above
receive?
Score:
\end{lstlisting}

\subsection{Additional Source Quality Analysis}
\label{appendix:source-quality-statistics}

\begin{figure*}[ht]
   \centering
    \includegraphics[width=1.0\textwidth]{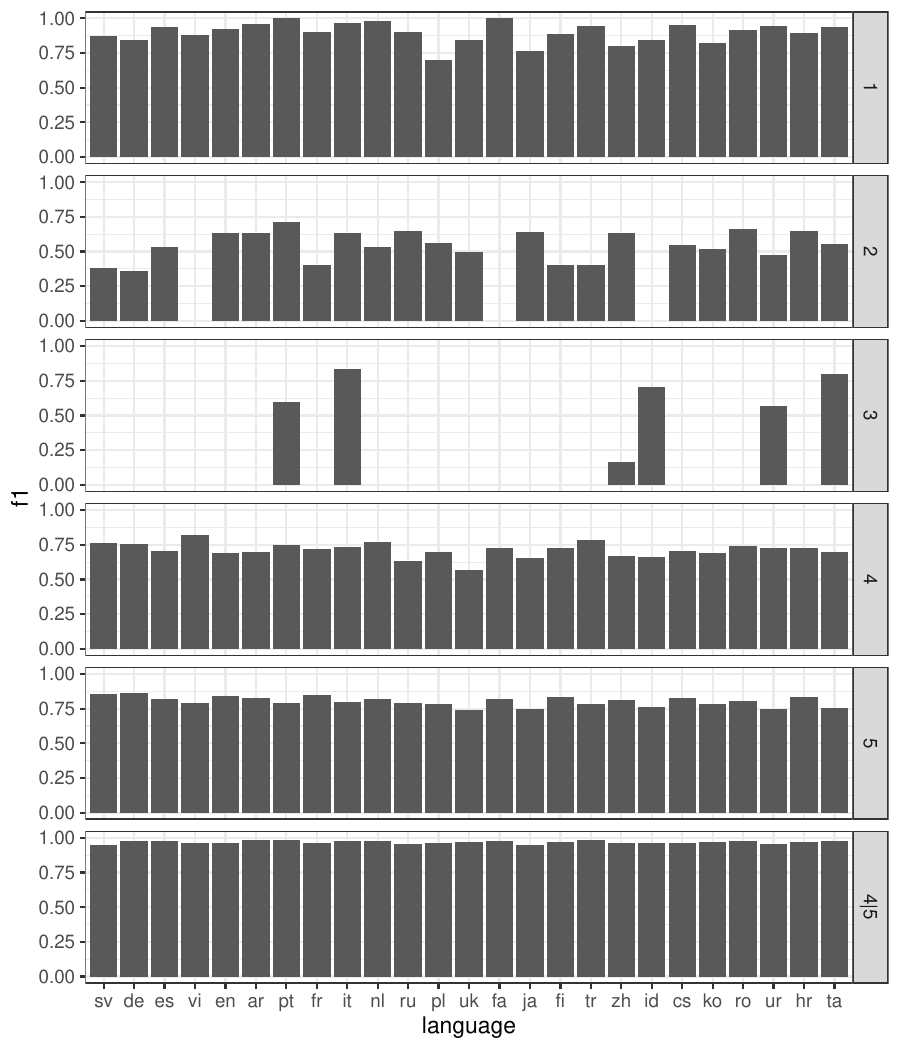}
    \caption{Per-class, per-language source quality prediction F1 scores of the XLM-RoBERTa regression model on the silver test set, for languages 1--25.  4|5 denotes the pseudo-label consisting of labels 4 and 5 combined, representing the set of ``higher-quality sources.''}
    \label{fig:source-quality-f1-per-language-1}
\end{figure*}

\begin{figure*}[ht]
   \centering
    \includegraphics[width=1.0\textwidth]{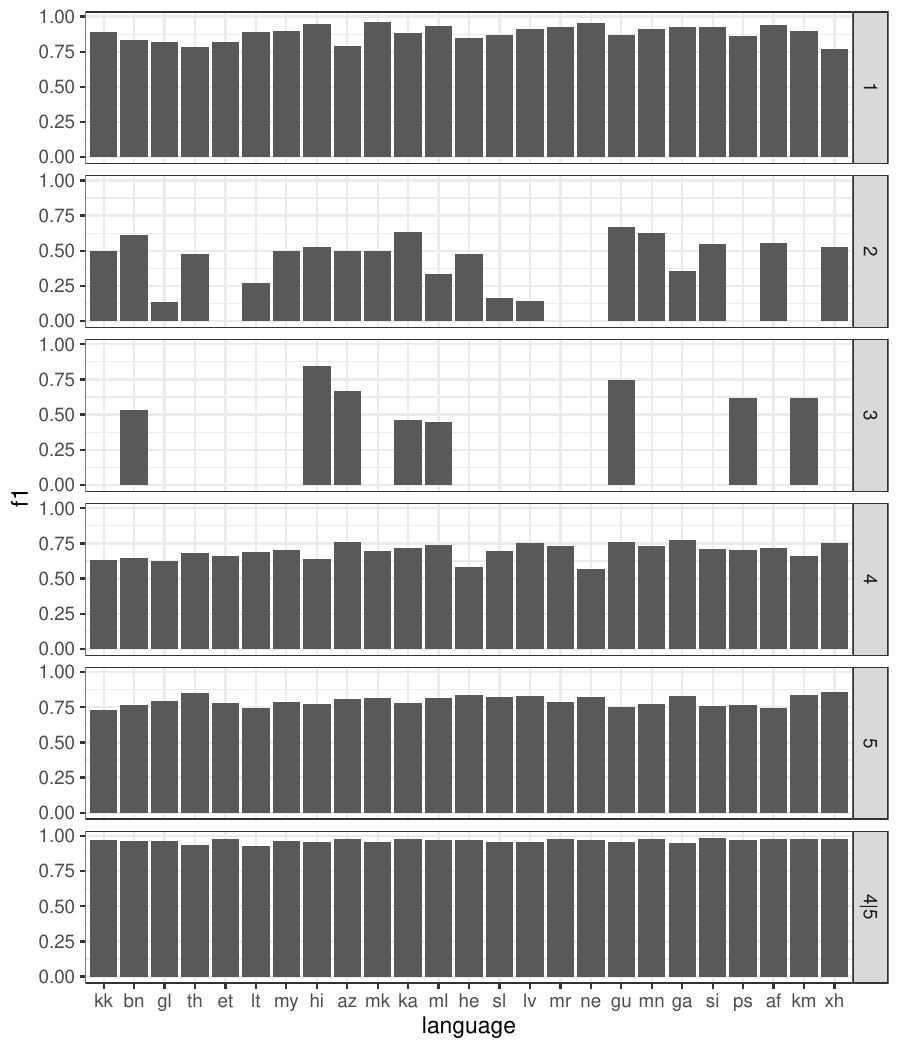}
    \caption{Per-class, per-language source quality prediction F1 scores of the XLM-RoBERTa regression model on the silver test set, for languages 26--50.  4|5 denotes the pseudo-label consisting of labels 4 and 5 combined, representing the set of ``higher-quality sources.''}
    \label{fig:source-quality-f1-per-language-2}
\end{figure*}

\begin{figure*}[ht]
   \centering
    \includegraphics[width=1.0\textwidth]{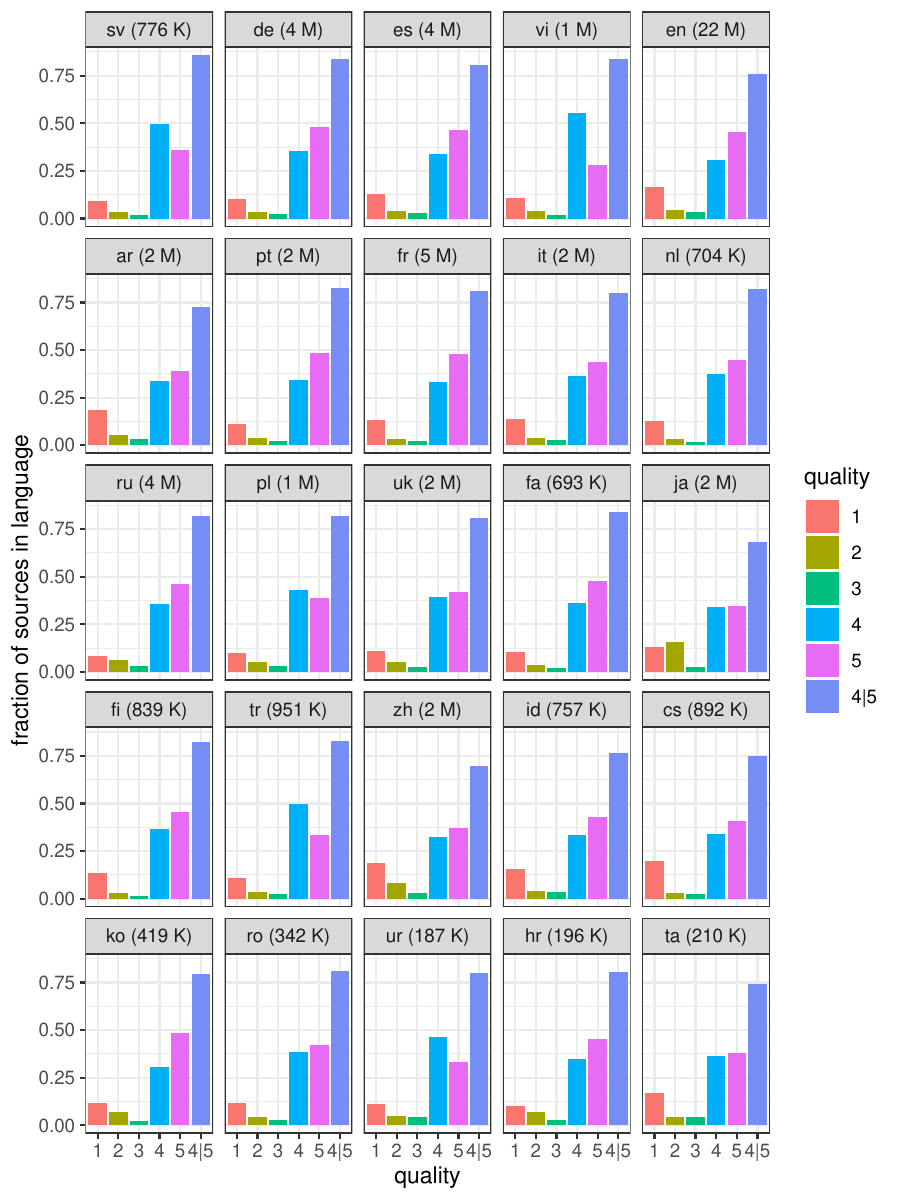}
    \caption{Source quality label distribution for languages 1--25.  4|5 denotes the pseudo-label consisting of labels 4 and 5 combined, representing the set of ``higher-quality sources.''  The number at the top of each panel is the total number of sources extracted in that language, where K is thousands and M is millions.}
    \label{fig:source-quality-per-language-1}
\end{figure*}

\begin{figure*}[ht]
   \centering
    \includegraphics[width=1.0\textwidth]{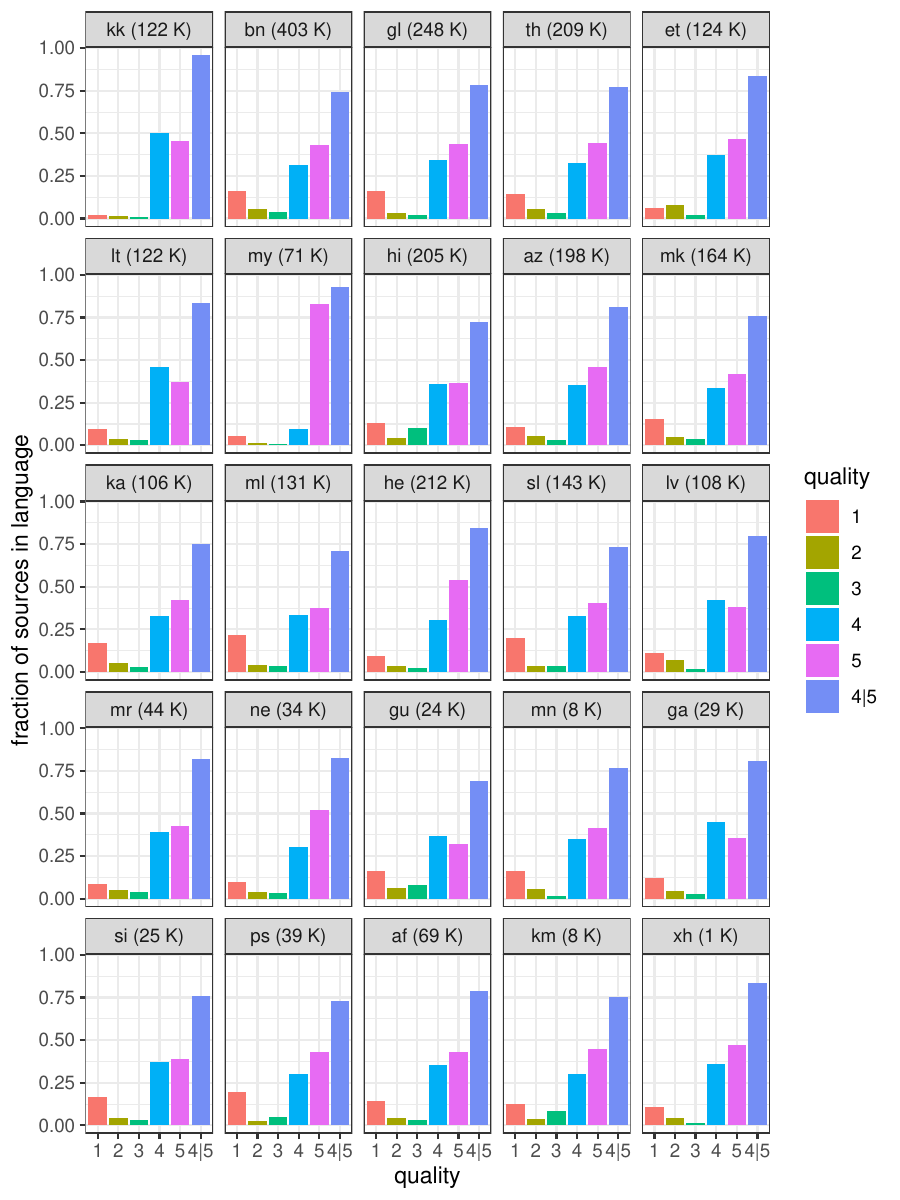}
    \caption{Source quality label distribution for languages 26--50.  4|5 denotes the pseudo-label consisting of labels 4 and 5 combined, representing the set of ``higher-quality sources.''  The number at the top of each panel is the total number of sources extracted in that language, where K is thousands and M is millions.}
    \label{fig:source-quality-per-language-2}
\end{figure*}

Additional plots and tables describing the source quality data and model for \megawika~2:
\begin{itemize}
\item Precision and recall of source quality models on the silver test set, including averages across language and measures of spread (Table~\ref{tab:source-quality-results-precision} and Table~\ref{tab:source-quality-results-recall}, respectively);
\item F1 of selected XLM-RoBERTa regression model for each class and language (Figure~\ref{fig:source-quality-f1-per-language-1} and Figure~\ref{fig:source-quality-f1-per-language-2});
\item Predicted source quality distributions for each language across all sources in \megawika~2 (Figure~\ref{fig:source-quality-per-language-1} and Figure~\ref{fig:source-quality-per-language-2}).
\end{itemize}
See Section~\ref{sec:source-text-quality-analysis} for main source text quality analysis.

\begin{table*}[ht]
\centering
\begin{tabular}{lrrrrrr}
\toprule
\textbf{Model} & \textbf{Class 1} & \textbf{Class 2} & \textbf{Class 3} & \textbf{Class 4} & \textbf{Class 5} & \textbf{Class 4|5} \\
\midrule
& \multicolumn{6}{c}{\emph{Average}} \\
\midrule
e5-small                             & 0.90 & 0.54 & 0.78 & 0.74 & 0.76 & 0.97 \\
e5-base                              & 0.90 & 0.56 & 0.79 & 0.70 & 0.79 & 0.97 \\
xlmr-base                            & 0.90 & 0.58 & 0.76 & 0.72 & 0.78 & 0.97 \\
\textsuperscript{*}xlmr-base regress & 0.91 & 0.48 & 0.58 & 0.73 & 0.80 & 0.98 \\
\midrule
& \multicolumn{6}{c}{\emph{Standard Deviation}} \\
\midrule
e5-small                             & 0.06 & 0.16 & 0.16 & 0.06 & 0.05 & 0.02 \\
e5-base                              & 0.07 & 0.21 & 0.26 & 0.06 & 0.06 & 0.02 \\
xlmr-base                            & 0.07 & 0.22 & 0.25 & 0.06 & 0.06 & 0.02 \\
\textsuperscript{*}xlmr-base regress & 0.07 & 0.14 & 0.18 & 0.06 & 0.05 & 0.01 \\
\midrule
& \multicolumn{6}{c}{\emph{Range}} \\
\midrule
e5-small                             & 0.25 & 0.83 & 0.50 & 0.25 & 0.23 & 0.09 \\
e5-base                              & 0.24 & 1.00 & 1.00 & 0.24 & 0.27 & 0.08 \\
xlmr-base                            & 0.23 & 0.82 & 1.00 & 0.28 & 0.25 & 0.09 \\
\textsuperscript{*}xlmr-base regress & 0.25 & 0.64 & 0.67 & 0.29 & 0.24 & 0.06 \\
\bottomrule
\end{tabular}
\caption{Average (macro), bias-corrected standard deviation, and range (max $-$ min) of \textbf{precision} across languages with at least five test examples on the silver test set.  4|5 is the pseudo-class consisting of the union of classes 4 and 5.  \textsuperscript{*}selected source quality model.}
\label{tab:source-quality-results-precision}
\end{table*}

\begin{table*}[ht]
\centering
\begin{tabular}{lrrrrrr}
\toprule
\textbf{Model} & \textbf{Class 1} & \textbf{Class 2} & \textbf{Class 3} & \textbf{Class 4} & \textbf{Class 5} & \textbf{Class 4|5} \\
\midrule
& \multicolumn{6}{c}{\emph{Average}} \\
\midrule
e5-small                             & 0.90 & 0.54 & 0.75 & 0.64 & 0.85 & 0.97 \\
e5-base                              & 0.88 & 0.49 & 0.67 & 0.72 & 0.78 & 0.97 \\
xlmr-base                            & 0.90 & 0.50 & 0.64 & 0.68 & 0.82 & 0.97 \\
\textsuperscript{*}xlmr-base regress & 0.88 & 0.55 & 0.69 & 0.68 & 0.81 & 0.96 \\
\midrule
& \multicolumn{6}{c}{\emph{Standard Deviation}} \\
\midrule
e5-small                             & 0.08 & 0.19 & 0.24 & 0.08 & 0.04 & 0.01 \\
e5-base                              & 0.08 & 0.19 & 0.27 & 0.07 & 0.06 & 0.01 \\
xlmr-base                            & 0.07 & 0.20 & 0.25 & 0.07 & 0.05 & 0.01 \\
\textsuperscript{*}xlmr-base regress & 0.08 & 0.19 & 0.25 & 0.07 & 0.05 & 0.02 \\
\midrule
& \multicolumn{6}{c}{\emph{Range}} \\
\midrule
e5-small                             & 0.35 & 0.77 & 0.67 & 0.35 & 0.23 & 0.05 \\
e5-base                              & 0.39 & 0.89 & 1.00 & 0.34 & 0.26 & 0.06 \\
xlmr-base                            & 0.35 & 0.75 & 1.00 & 0.31 & 0.22 & 0.06 \\
\textsuperscript{*}xlmr-base regress & 0.35 & 0.72 & 0.83 & 0.31 & 0.20 & 0.07 \\
\bottomrule
\end{tabular}
\caption{Average (macro), bias-corrected standard deviation, and range (max $-$ min) of \textbf{recall} across languages with at least five test examples on the silver test set.  4|5 is the pseudo-class consisting of the union of classes 4 and 5.  \textsuperscript{*}selected source quality model.}
\label{tab:source-quality-results-recall}
\end{table*}

\section{Translation}
\label{appendix:translation-extra}

\subsection{Sampling Implementation}

To sample translated passages (hereafter, for simplicity, \emph{passages}) from each dataset and model, for each language, we first uniformly sample a chunk from the first 30\% of chunks in the dataset, or 10 chunks, whichever is larger.  Then, for \megawika~1 (M2M and Google Translate), we filter out articles containing a colon or forward slash in the title (a proxy for special pages like talk pages and templates) and filter articles with less than a given minimum number of passages,\footnote{
    The minimum passage threshold varies per language; across languages, the threshold has minimum 2, Q1 6, median 8, Q3 10, and max 15.
}
and collect all passages from the remaining articles in the chunk.  Next, we filter to passage lengths between \num{15} and \num{2500} (inclusive) and filter out passages containing a colon or the terms ``\texttt{category}'' or ``\texttt{category}'' (case-insensitive),\footnote{
    The exception is Russian (ru), for which this passage filtering is not performed.
}
assign passages weights using to the weighting function in Section~\ref{sec:translation-analysis}, and sample a single passage.  We repeat that process, sampling with replacement, until we have \num{500} passages.

To sample passages for \megawika~2 (NLLB), we first sample a chunk as before.  We then uniformly sample an article from that chunk, and we form a single passage for each paragraph in that article by: filtering out sentences containing a colon or the terms ``\texttt{category}'' or ``\texttt{category}'', sampling a sentence count between 1 and 7 (uniformly), truncating to that many sentences, using the concatenation of the remaining sentences as the passages, and filtering to passage lengths between \num{15} and \num{2500} (inclusive) as before.  Finally, as before, we assign weights to those passages and sample a single passage.  We repeat the process, sampling with replacement, until we have \num{500} passages.

For \megawika~2 (NLLB), in the inner loop, we sample from each paragraph in an article rather than each passage in a chunk.  We also form passages from all sentences in a paragraph (not just those with citations), passages may contain non-contiguous sentences, and the sentences are taken from the beginning of a paragraph.\footnote{
    One might hypothesize that in many cases, early sentences make more general claims while later sentences make more specific claims, grounded in sources, to support the earlier, more general claims.  This structure resembles the standard ``3-3-3'' essay format, but at a smaller (paragraph) level.
}
Finally, chunks in \megawika~1 have variable length up to \num{500} articles, while all chunks but the last one in each language in \megawika~2 have \num{1000} articles.
Thus, passages sampled for NLLB are not \emph{directly} comparable to those sampled for M2M and Google Translate, even when controlling for differences in filtering and preprocessing between \megawika~1 and \megawika~2.

Due to an implementation error, $L_{\text{target}} = 160$ for M2M and Google Translate while $L_{\text{target}} = 150$ for NLLB.

All this said, the purpose of sampling is ultimately to produce similar collections of passages to compare, in particular, to produce collections of passages similar in length.  As reported in Table~\ref{tab:length_stats}, the \megawika~2 (NLLB) samples have wider spread, lower median, and intermediate mean compared to the \megawika~1 (M2M and Google Translate) samples, so we don't specifically anticipate a bias in favor of \megawika~2; we point out these discrepancies just to note there are additional forms of bias in the data.

\subsection{Passage Perplexity Distributions}

\begin{figure*}[ht]
   \centering
    \includegraphics[width=1.0\textwidth]{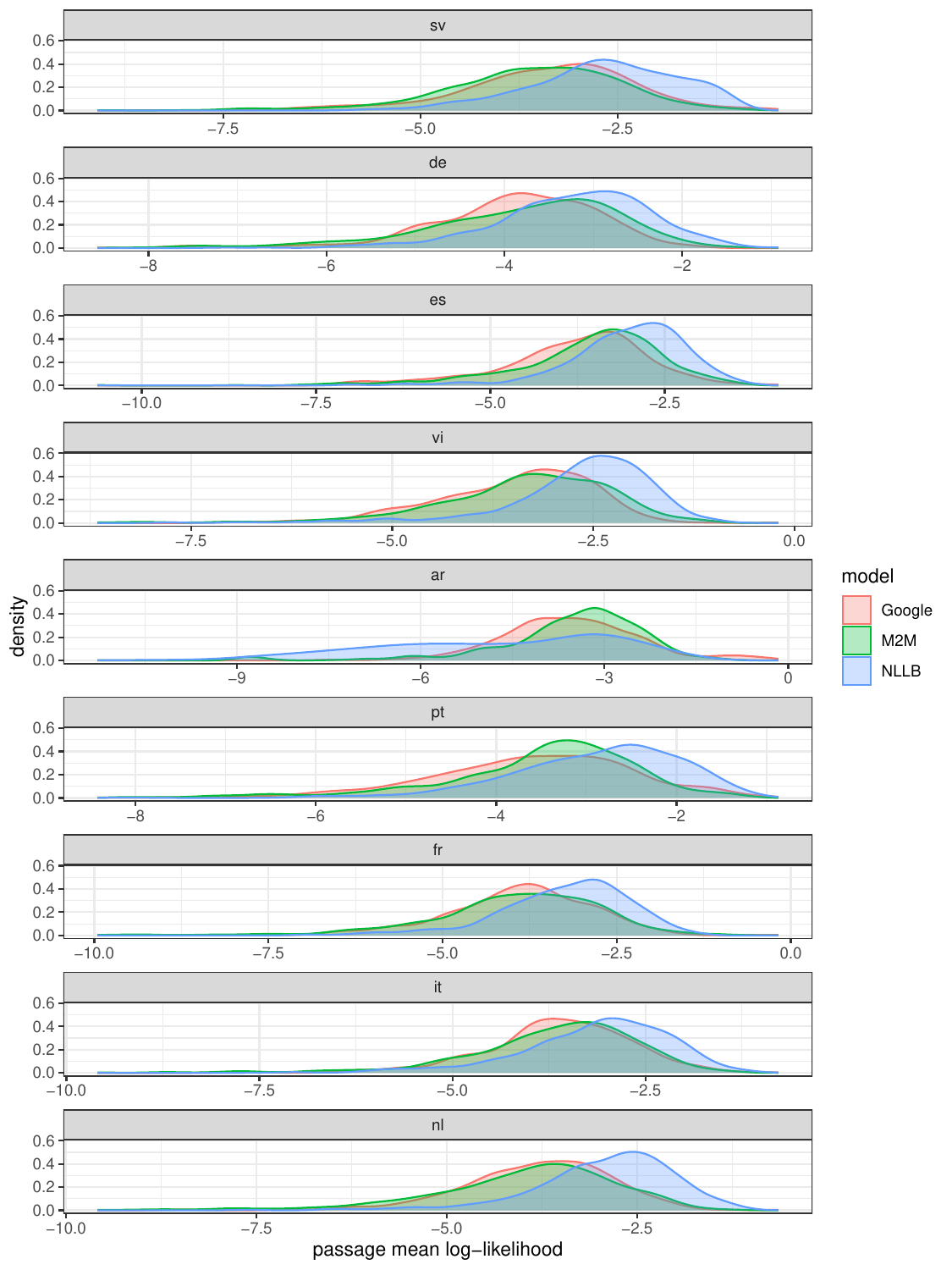}
    \caption{Sampled passage translation log-likelihood distribution for languages 1--10 (note English, en, is omitted).}
    \label{fig:translation-loglikelihood-density-1}
\end{figure*}

\begin{figure*}[ht]
   \centering
    \includegraphics[width=1.0\textwidth]{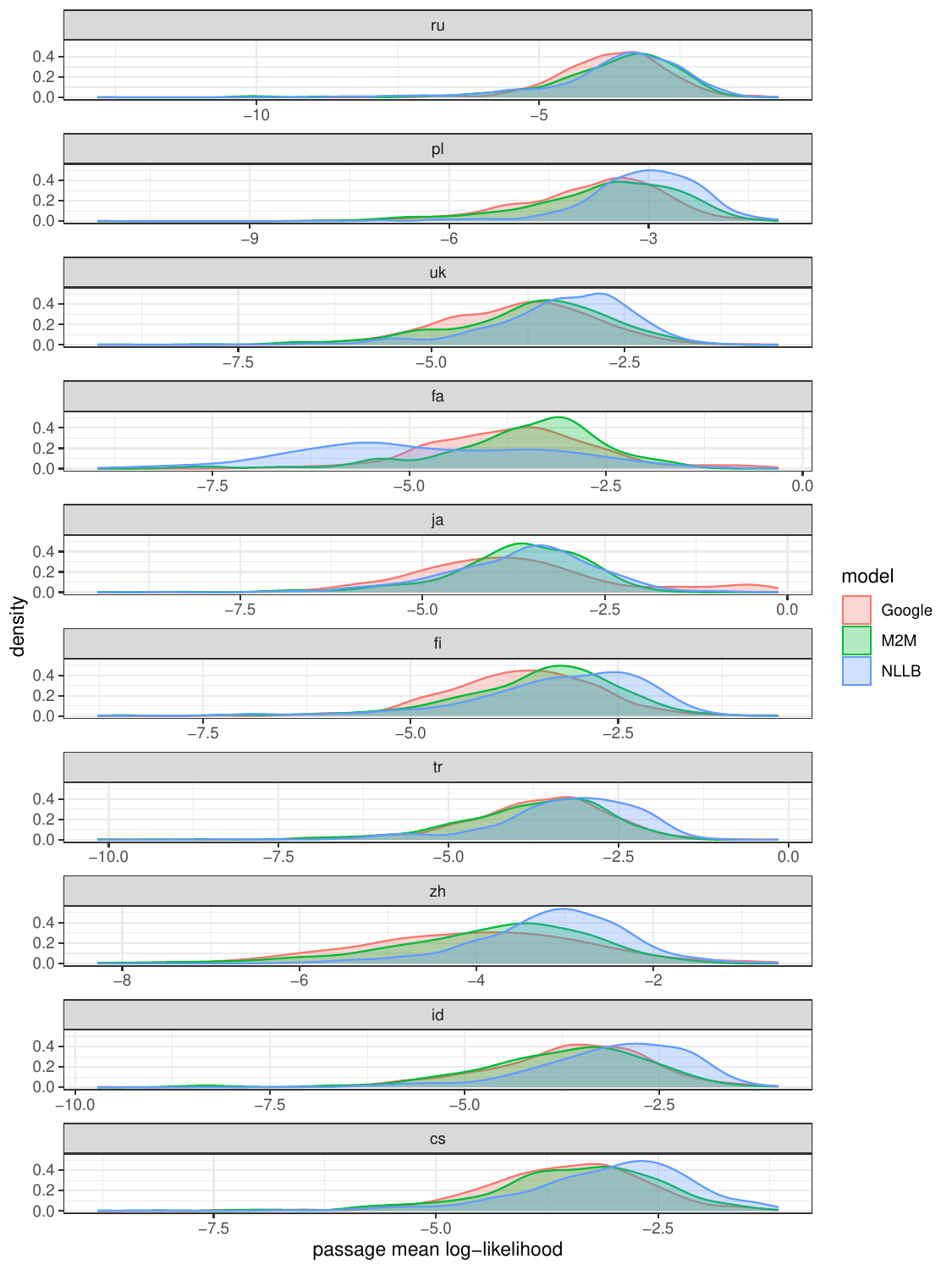}
    \caption{Sampled passage translation log-likelihood distribution for languages 11--20.}
    \label{fig:translation-loglikelihood-density-2}
\end{figure*}

\begin{figure*}[ht]
   \centering
    \includegraphics[width=1.0\textwidth]{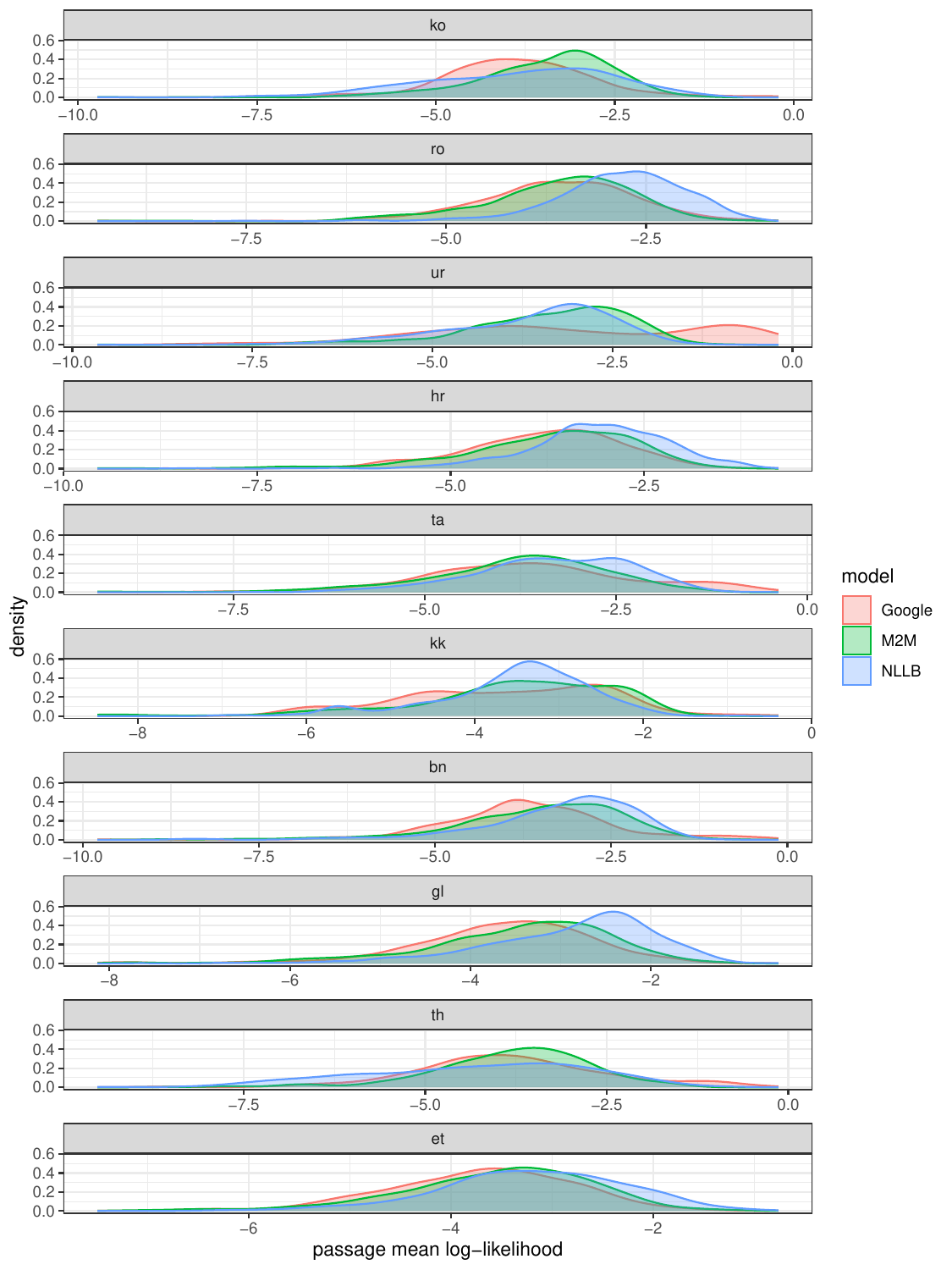}
    \caption{Sampled passage translation log-likelihood distribution for languages 21--30.}
    \label{fig:translation-loglikelihood-density-3}
\end{figure*}

\begin{figure*}[ht]
   \centering
    \includegraphics[width=1.0\textwidth]{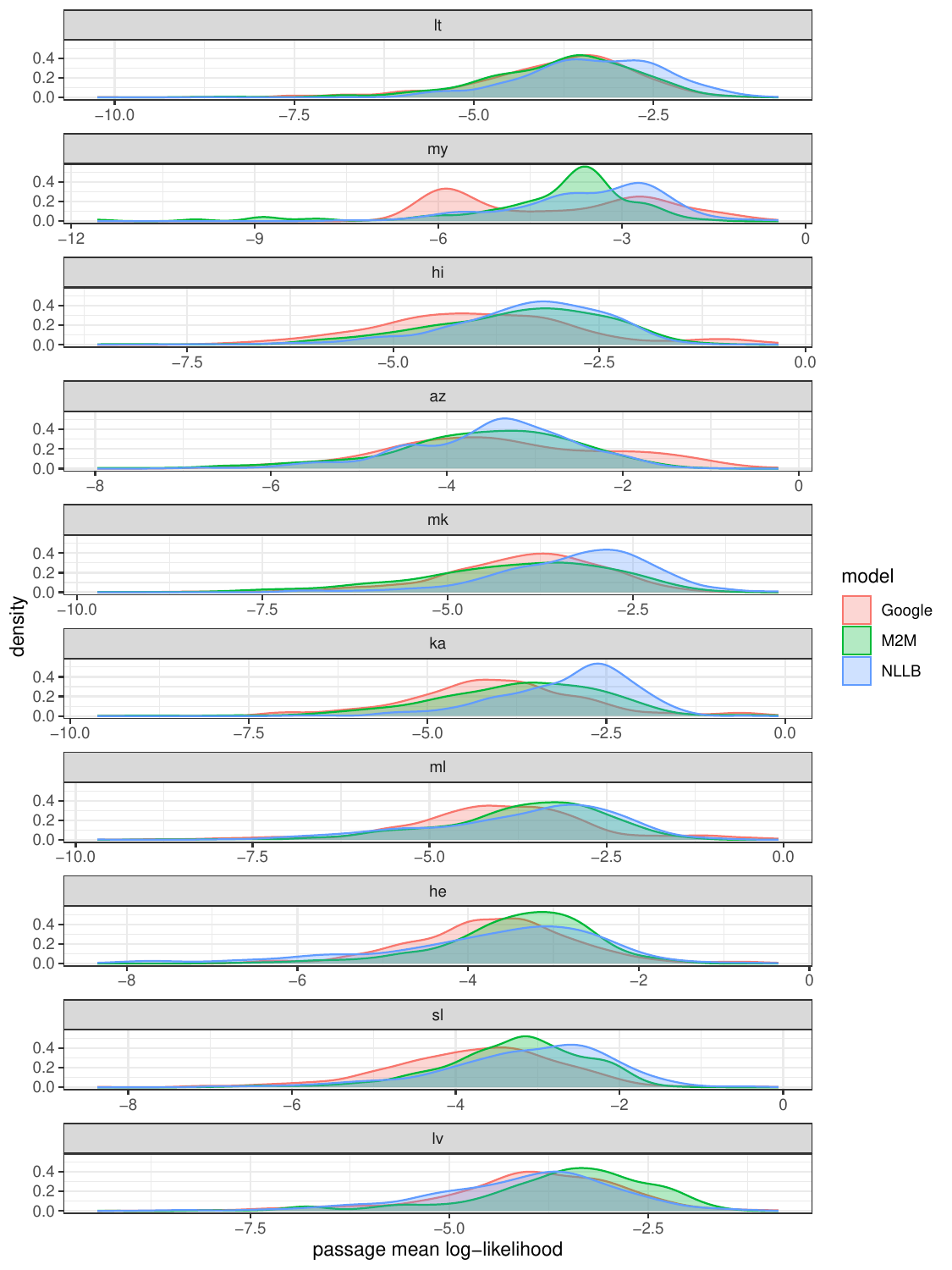}
    \caption{Sampled passage translation log-likelihood distribution for languages 31--40.}
    \label{fig:translation-loglikelihood-density-4}
\end{figure*}

\begin{figure*}[ht]
   \centering
    \includegraphics[width=1.0\textwidth]{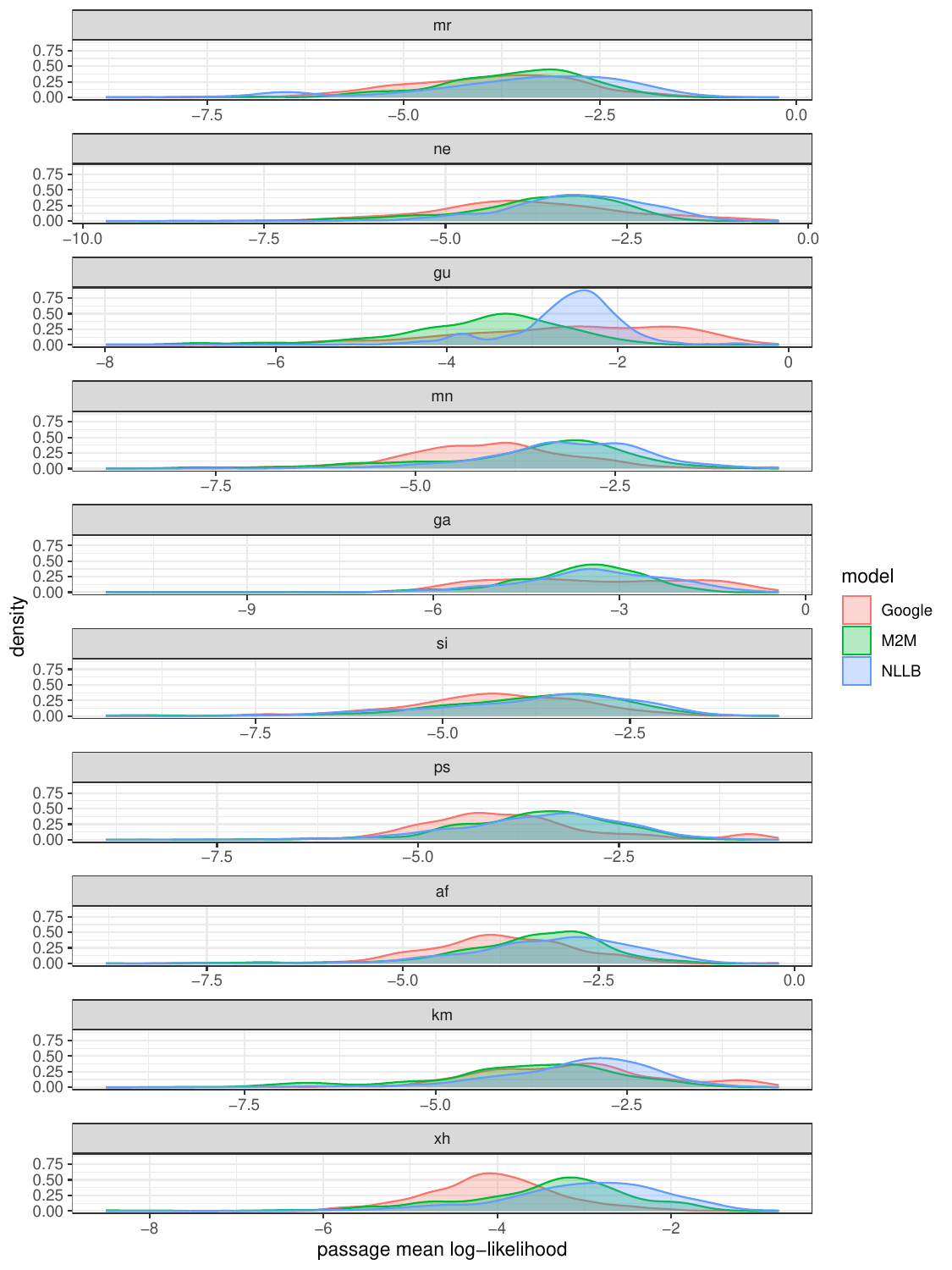}
    \caption{Sampled passage translation log-likelihood distribution for languages 41--50.}
    \label{fig:translation-loglikelihood-density-5}
\end{figure*}

Histograms of the sampled passage translation log-likelihoods for each language and model are provided in
Figures~\ref{fig:translation-loglikelihood-density-1},
\ref{fig:translation-loglikelihood-density-2},
\ref{fig:translation-loglikelihood-density-3},
\ref{fig:translation-loglikelihood-density-4},
\ref{fig:translation-loglikelihood-density-5}.  The distributions are split across multiple plots for clarity.  See Section~\ref{sec:translation-analysis} for the main translation analysis.

\section{WebBrain-Raw Sample Analysis}
\label{appendix:webbrain-sample-analysis}

WebBrain-Raw has 260 million references (web citations), many more than \megawika~2, despite spanning a fraction as many articles~\cite{qian2023webbrain}.  To investigate this discrepancy, we compared the 500-article WebBrain-raw sample against the same articles in \megawika~2.


We were able to find 447 articles in the sample with matching article titles in \megawika~2.  WebBrain-Raw had an average of 16 references per article on this subset while \megawika~2 had an average of 19 web citations per article.

We resolved page redirects on the remaining 53 article titles to test a hypothesis that there might be a larger disparity between \megawika~2 and WebBrain-Raw on this subset.  Resolving redirects using the Wikipedia Action API, we were able to locate articles for 47 of those remaining 53 titles.\footnote{
    The \megawika~2 Wikipedia dump was from May 2024, whereas we accessed the Action API in March 2025.  Additionally, redirects could have changed between the creation of WebBrain-Raw and the creation of \megawika~2 in ways that impact the results.
}
On the pages found by redirect resolution, \megawika~2 had an average of 51 web citations per article while WebBrain-Raw had an average of 40 references per article.

Aggregating these results, across the 494 pages in the WebBrain-Raw sample whose titles either appeared directly in \megawika~2 or which could be located in \megawika~2 after redirect resolution, \megawika~2 has an average of 22 web citations per article while WebBrain-Raw has an average of 19 references per article.

If the WebBrain-Raw sample is a uniform-at-random sample of the full data set, the number of references per article should be similar between the sample and the full data set.  The WebBrain-Raw sample has an average of 18 references per article while the full WebBrain-Raw data set has an average of 17 references per article, so the provided sample is consistent with uniform-at-random sampling.

\end{document}